\documentclass{aa}

\usepackage{natbib}
\usepackage{graphicx}
\usepackage{subcaption}
\usepackage{txfonts}
\usepackage{adjustbox} 
\usepackage[colorlinks = true, allcolors = blue]{hyperref}

\providecommand{\parallax}{\ensuremath{\varpi}}

\begin{document}

\title{Hunting for open clusters in \textit{Gaia} EDR3: $628$ new open clusters found with \texttt{OCfinder}\thanks{Full Table 1 and Table 2 are only available at the CDS.}}

\author{
    A. Castro-Ginard          \inst{\ref{inst:leiden},\ref{inst:UB}}\relax
\and   C. Jordi \inst{\ref{inst:UB}}\relax
\and   X. Luri  \inst{\ref{inst:UB}}\relax
\and   T. Cantat-Gaudin  \inst{\ref{inst:UB},\ref{inst:MPIA}}\relax
\and   J.M. Carrasco \inst{\ref{inst:UB}}\relax
\and   L. Casamiquela \inst{\ref{inst:bordeaux}}\relax
\and   F. Anders  \inst{\ref{inst:UB}}\relax
\and   L. Balaguer-N\'u\~{n}ez \inst{\ref{inst:UB}}\relax
\and   R.M. Badia \inst{\ref{inst:bsc}}
}

\institute{Leiden Observatory, Leiden University, Niels Bohrweg 2, 2333 CA Leiden, Netherlands\\ \email{acastro@strw.leidenuniv.nl}\relax \label{inst:leiden}
\and
Dept. F\'isica Qu\`antica i Astrof\'isica, Institut de Ci\`encies del Cosmos (ICCUB), Universitat de Barcelona (IEEC-UB), Mart\'i i Franqu\`es 1, E08028 Barcelona, Spain\relax \label{inst:UB}
\and
Max-Planck-Institut f\"ur Astronomie, K\"onigstuhl 17, D-69117 Heidelberg, Germany\relax \label{inst:MPIA}
\and
Laboratoire d'Astrophysique de Bordeaux, Univ. Bordeaux, CNRS, B18N, all\'ee Geoffroy Saint-Hilaire, 33615 Pessac, France\relax \label{inst:bordeaux}
\and 
Barcelona Supercomputing Center (BSC), Barcelona, Spain\relax \label{inst:bsc}}

\date{Received date /
Accepted date}

\abstract{
        The improvements in the precision of the published data in \textit{Gaia} EDR3 with respect to \textit{Gaia} DR2, particularly for parallaxes and proper motions, offer the opportunity to increase the number of known open clusters in the Milky Way by detecting farther and fainter objects that have thus far gone unnoticed.
}{
        Our aim is to continue to complete the open cluster census in the Milky Way with the detection of new stellar groups in the Galactic disc. We use \textit{Gaia} EDR3 up to magnitude $G = 18$ mag, increasing the magnitude limit and therefore the search volume explored in one
unit in our previous studies.
}{
        We used the \texttt{OCfinder} method to search for new open clusters in \textit{Gaia} EDR3 using a big data environment. As a first step, \texttt{OCfinder} identified stellar statistical overdensities in five-dimensional astrometric space (position, parallax, and proper motions) using the \texttt{DBSCAN} clustering algorithm. Then, these overdensities were classified into random statistical overdensities or real physical open clusters using a deep artificial neural network trained on well-characterised $G$, $G_{\rm BP} - G_{\rm RP}$ colour-magnitude diagrams.
}{
        We report the discovery of $628$ new open clusters within the Galactic disc, with most of them being located beyond $1$ kpc from the Sun. From the estimation of ages, distances, and line-of-sight extinctions of these open clusters, we see that young clusters align following the Galactic spiral arms while older ones are dispersed in the Galactic disc. Furthermore, we find that most open clusters are located at low Galactic altitudes with the exception of a few groups older than $1$ Gyr. 
}{
        We show the success of the \texttt{OCfinder} method leading to the discovery of a total of $1\,274$ open clusters (joining the discoveries here with the previous ones based on \textit{Gaia} DR2), which represents almost $50\%$ of the known population. Our ability to perform big data searches on a large volume of the Galactic disc, together with the higher precision in \textit{Gaia} EDR3, enable us to keep completing the census with the discovery of new open clusters.
}
\maketitle


\section{Introduction}
\label{sec:intro}

Open clusters (OCs) have historically been used to study the structural, kinematical, and chemical properties of the disc of the Milky Way, and its evolution. For this reason, in recent years, there has been a growing interest in building an accurate and complete view of the OC population. In particular, after the \textit{Gaia} Second Data Release \citep[\textit{Gaia} DR2,][]{2018A&A...616A...1G}, the study of this field was revolutionised by the re-definition of the OC population, with \citet{tristan_catalogue} refusing around $50\%$ of the OCs reported in pre-\textit{Gaia} catalogues as not real OCs \citep{dias,kharchenko}. Furthermore, \textit{Gaia} DR2 enabled the systematic detection of new OCs using machine-learning methods, which outperform traditional manual methods to search for these objects. \citet[hereafter PaperI]{acastro1} were able to find $23$ new OCs in \textit{Gaia} DR1's TGAS subset \citep{2016A&A...595A...2G,tgas}, and then they used the same methodology to detect more than $600$ new OCs in the Galactic disc using \textit{Gaia} DR2, which represents about one-third of the currently known OC population \citep[hereafter PaperII and PaperIII, respectively]{acastro2,2020A&A...635A..45C}. Since then, several publications have made use of machine-learning-based methods to detect new OCs \citep{coin_clusters,upk_clusters,2019ApJS..245...32L,2020MNRAS.496.2021F,2021A&A...646A.104H}, computing membership lists \citep{2020A&A...633A..99C,2021ApJ...923..129J} or characterising their astrophysical properties \citep{2019A&A...623A.108B,2020A&A...640A...1C,2021MNRAS.504..356D}.

Several studies have combined \textit{Gaia} astrometry with ground-based radial velocities with the purpose of studying the kinematics of the OC population \citep[e.g.][]{2018A&A...619A.155S,2022A&A...658A..14C}. \citet{2021A&A...647A..19T} studied the 3D kinematics and age dependence of the OC population, also providing orbital parameters for $1\,382$ OCs. Other studies have used OCs' available astrometric and kinematic information to trace the spiral structure in the Milky Way \citep{2005ApJ...629..825D}. On this topic, \citet{2021FrASS...8...62M} found that the behaviour of the spiral arms could be explained by classical density waves. However, \citet{2021A&A...652A.162C} recently found a transient nature of the arms disfavouring classic density waves as the main drivers of the spiral structure, which is also supported by studies using tracers other than OCs \citep{2021A&A...654A.138M,2022A&A...658A..54C}. These contradicting results show the need to keep improving the OC census.

The coupling of \textit{Gaia} data with the detailed abundance results of large spectroscopic surveys has allowed for a more complete picture of the chemical composition of the OC population to be sketched. In all of these studies, an unbiased and complete census of the OCs of the Milky Way is needed to tackle the chemical evolution of our Galaxy. Detailed chemical abundance radial gradients in the Milky Way and their age dependence have been characterised \citep[e.g.][]{2019A&A...623A..80C,2021MNRAS.503.3279S}. Temporal dependencies of chemical abundance ratios are commonly calibrated using OCs due to their precise age determination \citep{2021A&A...652A..25C}. Finally, OCs are used as test cases to explore the feasibility of a diversity of techniques such as strong chemical tagging, that is to say the possibility of finding stars that were born in the same star-forming event \citep{2021A&A...654A.151C}.

The latest release of \textit{Gaia} data \citep[EDR3,][]{2021A&A...649A...1G}, providing astrometric measurements for about $1.8$ billion stars with improved precision with respect to \textit{Gaia} DR2, offers the opportunity to re-visit the OC census and keep improving it, both in terms of a better characterisation and new discoveries. The improved precisions in parallax, and particularly in proper motions with respect to \textit{Gaia} DR2, allow for the application of machine-learning methods to search for new structures that would go unnoticed with traditional methods, which mostly relied on visual inspection. In this context, we adapted our methodology (developed and applied in Paper I, Paper II, and Paper III), which we dub \texttt{OCfinder}, to search for unknown OCs in \textit{Gaia} EDR3 using its astrometric and photometric data.

This paper is organised as follows. In Sect.~\ref{sec:data}, we describe the data used to search for OCs. The \texttt{OCfinder} method used for that purpose is described in Sect.~\ref{sec:method}. Section~\ref{sec:results} describes the OCs found, both re-detected and new findings. Finally, our conclusions are presented in Sect.~\ref{sec:conclusions}.
\section{Data}
\label{sec:data}

The data were used in this paper to search for unknown OCs is \textit{Gaia} EDR3 \citep{2021A&A...649A...1G}. \textit{Gaia} EDR3 is the first delivery of the third data release, and among other products it provides about $1.8$ billion sources with astrometric and photometric observations, that is $(l,b,\varpi,\mu_{\alpha^*},\mu_\delta,G,G_{\rm BP},G_{\rm RP})$, with an improved precision with respect to \textit{Gaia} DR2 due to an increase in the observational time baseline, now spanning a period of $34$ months. Thanks to this improvement, we are able to search for new OCs with a deeper magnitude cut, which allows us to reach farther and less populated groupings. The adopted magnitude limit is $G = 18$ mag, unlike our previous studies where the limit was $G = 17$. Additionally, and since OCs are usually found in the Galactic disc, we limited our search to Galactic latitudes within $|b| \leq 20^{\circ}$, where most of the OCs are expected to be found. Similarly to our previous searches, we rejected stars with parallaxes larger than $7$ mas to avoid very close OCs, which will suffer from strong projection effects and will not be detectable by our method, and sources with negative parallaxes. We also filtered out stars with proper motions higher than $|\mu_{\alpha^*}|$ and $|\mu_{\delta}| \geq 30$ mas yr$^{-1}$ to remove stars incompatible with disc rotation, which OCs are expected to follow.

The median parallax uncertainty in \textit{Gaia} EDR3 at $G = 18$ mag is $0.12$ mas, while for proper motions the median uncertainties in $\mu_{\alpha^*}$ and $\mu_\delta$ at $G = 18$ mag are of $0.123$ and $0.111$ mas yr$^{-1}$, respectively \citep{2021A&A...649A...2L}. These are similar uncertainty levels in comparison to that of \textit{Gaia} DR2 at magnitude $G = 17$ mag \citep{2018A&A...616A...2L}.  For the photometry, the uncertainties in \textit{Gaia} EDR3 at $G = 18$ mag are at the level of a thousandth for $G$, and a hundredth of a magnitude for $G_{\rm BP}$ and $G_{\rm RP}$ \citep{2021A&A...649A...3R}. These magnitude uncertainty levels are also comparable to \textit{Gaia} DR2 at $G = 17$ mag; therefore we consider them to be a reasonable compromise to succesfully achieve our goals. Altogether, and taking the aforementioned filters into account, the sample to be analysed contains $232\,463\,114$ sources, and its analysis is enabled thanks to the use of a big data environment in our data analysis pipeline \citep{2020A&A...635A..45C}.

\section{The \texttt{OCfinder} method}
\label{sec:method}

The methodology developed to search for new OCs in \textit{Gaia} data, \texttt{OCfinder}, is described in detail in Paper I. It was successfully applied to detect 23 new nearby OCs \citep{acastro1} in the TGAS data set of \textit{Gaia} DR1. It was also applied to \textit{Gaia} DR2 where $53$ new OCs were detected in a direction near the Galactic anticentre \citep{acastro2} and hundreds of new OCs in a big data search on the whole Galactic disc \citep{2020A&A...635A..45C}.

The \texttt{OCfinder} method consists of two main steps. The first step is a blind search for overdensities in the five-dimensional astrometric space of \textit{Gaia}, that is $(l,b,\varpi,\mu_{\alpha^*},\mu_\delta)$, so as to find sets of stars which are more clustered than the average field stars for that region (Sect.~\ref{subsec:clustering}). The second step makes use of the \textit{Gaia} photometry to confirm whether OC member stars follow an isochrone pattern in a colour-magnitude diagram (CMD) using an artificial neural network trained on well-characterised CMDs (Sect.~\ref{subsec:ann}).

\subsection{Data preparation}
\label{subsec:preprocessing}

We divided the sky into small areas of size $L\times L$ deg$^2$, where $L$ varies according to the local structure density. This was done in order to define local average densities, accounting for the varying densities of the Galactic disc, when searching for representative overdensities which may correspond to physical OCs. In our methodology, the sizes of these regions are not defined by following computational, but physical arguments. We used the \textit{Gaia} Universe Model Snapshot\footnote{GUMS (true values of instrinsic simulated sources) and GOG (observed attributes with simulated observational uncertainties) can be found in the \textit{Gaia} archive: \url{https://gea.esac.esa.int/archive/}} \citep[GUMS,][]{2012A&A...543A.100R} to represent the field star population, together with realistic OCs simulated using the \textit{Gaia} Object Generator \citep[GOG,][]{gog} both including errors at the time of \textit{Gaia} EDR3\footnote{Computed with the prescription given in \url{https://github.com/agabrown/PyGaia}}, to find the size $L$ of the regions that detect most of the simulated OCs (see Sect.~3 in PaperI for details). In this case, the sizes of the regions ranges from $L = 10^\circ$ to $L = 15^\circ$, which corresponds to a maximum of about $10^7$ stars per box to be simultaneously analysed with the clustering algorithm. The simultaneous analysis of such a large number of stars is not a problem for our method due to the inclusion of a big data environment (see details in Sect~\ref{subsec:clustering}).

Once the stars are divided into the $L\times L$ deg$^2$ regions, the five astrometric dimensions $(l,b,\varpi,\mu_{\alpha^*},\mu_\delta)$ are standardised in order to balance their importance in the clustering algorithm. In our case, we used the \texttt{StandardScaler} method implemented in the \texttt{scikit-learn} Python library \citep{sklearn},  which transforms each dimension to have zero mean and unit variance. 

\subsection{Astrometric clustering with DBSCAN}
\label{subsec:clustering}

In each of the $L\times L$ deg$^2$ regions, we ran the density-based clustering algorithm \texttt{DBSCAN} \citep{dbscan} to find statistical overdensities that may belong to real OCs. \texttt{DBSCAN} relies on two input parameters, which are $\epsilon$ and $minPts$, to define a density threshold and it searches for overdensities above the threshold. As a brief description, \texttt{DBSCAN} visits each source in the dataset, builds an $N$-dimensional $\epsilon$ neighbourhood around the source, and counts how many sources are within the $\epsilon$ neighbourhood. If at least $minPts$ sources are found, they are considered to be a cluster (we refer readers to Sect.~2 in Paper I for a \texttt{DBSCAN} description relevant for this application).

Similarly to our previous applications of \texttt{OCfinder}, we selected several optimal values of $minPts$ which were found, together with $L$, using simulated data. In this case, the values of $minPts$ range from eight to $16$ stars. The computation of the $\epsilon$ parameter was done automatically in each $L\times L$ deg$^2$ region taking advantage of the fact that OC member stars are closer than random field stars in multidimensional space including positions, parallax, and proper motions. In brief, we computed the distribution of distances for each star to its $k_{th}$ nearest neighbour (defined as $k = minPts - 1$), and we compared it to the distribution of $k_{th}$ nearest neighbour distances between field stars (with no substructure present). Then, $\epsilon$ is defined as the minimum $k_{th}$ distance between field stars, below where the distribution starts to differ from the real distribution (due to the presence of clusters in the latter). Again, we refer the reader to Sect.~2.2 in Paper I for exact details on the computation of $\epsilon$. 

After \texttt{DBSCAN} was applied to a given $L\times L$ deg$^2$ region, we shifted these regions by $L/3$ and $2L/3$ and applied \texttt{DBSCAN} again in the new region in order to account for clusters in the borders of the grid. Then, we merge our duplicated or ovelapping groupings that are in fact a single cluster. The whole process was applied using several values for the pairs $(L,minPts)$. This way, a Monte Carlo-like analysis of the results was enabled, and clusters with more findings within the different pairs of $(L,minPts)$ are the most reliable. 

The choice of \texttt{DBSCAN} is due to i) its ability to work with $N$-dimensional data, ii) the fact that it can handle noise (sources not assigned to any cluster), iii) being density-based, it can account for projections effects and clusters not having a predetermined shape, and iv) the fact that it does not require an a priori number of clusters to be found. The caveat of \texttt{DSBCAN} is that it is limited to a single density threshold (defined by $\epsilon$ and $minPts$), and it only finds overdensities above that limit. This has been improved with \texttt{HDBSCAN} \citep[Hierarchical-\texttt{DBSCAN},][]{hdbscan}, with which a whole range of $\epsilon$ values is explored, therefore allowing there to be clusters with different density thresholds. In this case, however, the number of clusters found drastically increases, thus increasing the number of false positives and the complexity in the interpretability of the results. We consider that in our approach, using \texttt{DBSCAN} with several pairs of $(L,minPts)$ which are found to be efficient in detecting a large number of clusters using realistic simulated data, we cover the range of densities which may define OCs. A comparison between different clustering algorithms, including \texttt{HDBSCAN}, \texttt{DBSCAN,} and our specific approach in \texttt{OCfinder}, was carried out by \citet{2021A&A...646A.104H}, who found \texttt{OCfinder} to be the best among the explored options in terms of balance in sensitivity, specificity, and precision.

The whole clustering process is deployed at the MareNostrum supercomputer, located at the Barcelona Supercomputing Center\footnote{\url{https://www.bsc.es/marenostrum}}. Each \texttt{DBSCAN} run for each $(L,minPts)$ pair was launched distributed in three nodes of MareNostrum (with a total of $144$ cores and $48$ cores for each node). To distribute the process, we used \texttt{PyCOMPSs} \citep{pycompss}, a Python-based application that distributes and schedules the execution of a job transparently to the user. Execution times for each \texttt{DBSCAN} application on the whole Galactic disc range from $12$ to $27$ hours depending on the $(L,minPts)$ pairs, with higher values for both $L$ and $minPts$ being more computationally expensive due to the larger amount of sources to analyse. The advantage of using \texttt{PyCOMPSs}, as well as the big data environment of MareNostrum, can be seen in \citet{dislib}, where the authors compare the performance of the clustering process of \texttt{OCfinder} in different environments.

\subsection{Photometric confirmation with deep learning}
\label{subsec:ann}

The second step of \texttt{OCfinder} is the recognition of physical OCs among the statistical clusters found by \texttt{DBSCAN}. We built CMDs from the members of each statistical cluster using \textit{Gaia}'s $G$, $G_{\rm BP}$, and $G_{\rm RP}$ photometry, and we used a deep artificial neural network \citep[ANN,][]{ann} trained on well-characterised CMDs to distinguish real OCs by detecting their characteristic isochrone patterns. In our first applications of \texttt{OCfinder}, in Paper I and Paper II, we used a multi-layer perceptron with a single hidden layer to perform the classification. The big data search in the whole Galactic disc in Paper III resulted in a larger amount of statistical clusters found (with respect to Paper I and Paper II). There, we used a more robust classification through a deep ANN architecture, which outperformed the simpler ANN.

The deep ANN consists on an initial set of convolutional layers that extract the meaningful characteristics of the CMDs, followed by a set of fully connected layers to perform the classification. We used the PyTorch\footnote{\url{https://pytorch.org/}} package \citep{pytorch}, which provides powerful software particularly well suited for deep learning, to implement our deep ANN. We used the CUDA environment \citep{cuda} to implement and train the deep ANN on a NVidia RTX 2080Ti GPU, which provides fast and agile computations that allowed us to test different ANN architectures until reaching the optimal configuration.

The key ingredient for a good classification result is to ensure a high-quality training set. We used the largest homogeneous sample of known OCs present in \citet{2020A&A...640A...1C} to represent positive isochrone identifications. From this list, we removed clusters with very few stars (at least $minPts$) up to magnitude $G = 18$ mag, clusters with diffuse isochrones, and highly contaminated cases. We used data augmentation techniques to increase the number of training examples, meaning that for each OC we built new CMDs from a subpopulation of the original OC members. We supplemented this training set with simulated clusters which were generated using synthetic isochrones assuming solar metallicity ($Z \simeq 0.0152$ dex) from the PARSEC code\footnote{\url{http://stev.oapd.inaf.it/cgi-bin/cmd}} \citep{2012MNRAS.427..127B}, with ages ranging from $4$ Myr to $13$ Gyr, approximately. For each synthetic population, we built different sub-samples using the same data augmentation techniques as in the case of the real OCs, which were placed at different distances ranging from $300$ pc to $4$ kpc, as well as different values for extinction $A_v$ ranging from $0$ to $2$ mag, in order to represent all the possible configurations in the CMD. In order to mimic \textit{Gaia} EDR3 photometric observations, we added photometric errors in each band (described in Appendix~\ref{app:photometry_error}). 

On the negative identification side (identification of no clusters), we used random field stars queried from the \textit{Gaia} EDR3 archive at locations that avoid known open clusters. We also applied our \texttt{DBSCAN} approach to the GUMS and used the resulting statistical clusters, which do not represent real objects since stellar groupings of such OCs are not present in GUMS, to increase the negative identification training set. Similarly to our previous studies, in order to feed the network and perform the classification, we converted each CMD to a $2D$ histogram to extract their pixels, which were then normalised. We used a logarithmic normalisation scheme in order to enhance the lower density regions which are key for the characterisation of contaminants in the CMD.

\section{Results}
\label{sec:results}

\subsection{Crossmatch to known cluster catalogues}
\label{subsec:comparison}

The OCs found with our \texttt{OCfinder} methodology were crossmatched to known OC catalogues to identify which OCs are already known and which are new findings. The source of most known identifications is the catalogue provided by \citet{2020A&A...640A...1C}, representing the largest homogeneous catalogue including $2\,017$ OCs with information about their mean astrometric parameters, as well as estimated values for age, distance, and line-of-sight extinction. We consider our OCs to match with a cluster in \citet{2020A&A...640A...1C} if their centres are within a circle of radius $0.5^\circ$ in $l$ and $b$ coordinates, and if their mean parallaxes and proper motions are compatible within $2\sigma$ (where $\sigma$ is the quadratic sum of the uncertainties quoted in both catalogues for each quantity). In the first step of \texttt{OCfinder}, the application of \texttt{DBSCAN} was able to find $1\,559$ clusters from \citet{2020A&A...640A...1C} which represents nearly $80\%$ of the catalogue. In the second \texttt{OCfinder} step, the photometric confirmation, the ANN validated $1\,515$ of the crossmatched OCs previously found with \texttt{DBSCAN}, which shows the high efficiency of the ANN in identifying OC CMDs against random statistical overdensities. This re-detection efficiency is similar to our previous work in Paper III using \textit{Gaia} DR2, and it is mostly due to the selection of hyper-parameters for \texttt{DBSCAN} that are optimised for an all-sky search (see Sect.~\ref{subsec:clustering}). 

Recently, \citet{2021MNRAS.504..356D} have provided fundamental parameters for $1\,743$ OCs in our Galaxy based on \textit{Gaia} DR2. The vast majority of these clusters are also included in \citet{2020A&A...640A...1C}, and therefore they have already been crossmatched. However, there are clusters which could not be characterised by \citet{2020A&A...640A...1C} or new clusters which were detected afterwards \citep[e.g.][]{2020MNRAS.499.1874M,2020MNRAS.496.2021F}. From them, we were able to re-detect $110$ in our clustering step of which $104$ were confirmed in the photometric validation using the CMD. \citet{2021MNRAS.504..356D} also provide a list of dubious and likely not real OCs. We could not detect any of the OCs listed as they are not likely real, thus confirming the results by \citet{2021MNRAS.504..356D} for these cases. On the other hand, we were able to re-detect eight out of $11$ UBC clusters detected in Paper III which are listed as dubious; these include UBC~$359$, UBC~$416$, UBC~$505$, UBC~$573$, UBC~$575$, UBC~$577$, UBC~$579$, and UBC~$593$. We were not able to re-detect any of the \citet{2019ApJS..245...32L} candidates listed as dubious.

In the aforementioned catalogues, there is a large contribution of UBC clusters detected in Paper I, Paper II, and Paper III. From the $\sim650$ UBC clusters, we were able to re-detect $514$ ($\simeq80\%$) of them with the \textit{Gaia} EDR3 data, which is a similar re-detection efficiency as in the general case. For the different releases of UBC clusters, our re-detection efficiency is $\simeq30\%$ for Paper I, $\simeq60\%$ for Paper II, and $\gtrsim80\%$ for Paper III. The differences in the detection efficiency for these UBC clusters are due to the different star densities, and nature, of the datasets analysed, captured in the hyper-parameters used for the search. In Paper I and Paper II, the OC search was performed on the TGAS subset of \textit{Gaia} DR1 (where the limiting magnitude is $G=12$ mag) and a low density region localised near the Galactic anticentre in \textit{Gaia} DR2, respectively. In those searches, the hyper-parameters for the \texttt{DBSCAN} were adapted to the corresponding regions, and they are different from the all-sky search performed in this work. For OCs in Paper III, the detection efficiency is slightly higher because of the similarity between both datasets (see Sect.~\ref{sec:data}), and thus this is also the case in the method hyper-parameters.

\citet{2021A&A...646A.104H} recently reported the discovery of $41$ new OCs using a \texttt{HDBSCAN} clustering algorithm on \textit{Gaia} DR2. Out of these, we were able to re-detect $20$ of them in \textit{Gaia} EDR3 with our \texttt{OCfinder} method. From the $21$ remaining clusters, nine were detected in the first \texttt{DBSCAN} step, but they were not validated in the second ANN step. Therefore, they need to be further investigated. The reason for the non-detection of the other $12$ clusters can be related to the choice of the algorithm, among other causes. In our \texttt{OCfinder} method, we used \texttt{DBSCAN} to detect astrometric overdensities which are limited to a single density threshold, usually limited to the densest cluster in the region (a  limitation we minimised by performing a Monte-Carlo-like analysis, see Sect.~\ref{subsec:clustering}). \texttt{HDBSCAN} runs the clustering in a hierarchy of density thresholds, thus it is able to detect varying density clusters on the same region, at the cost of increasing the number of false positives. This is seen in the compactness of the OCs found by \citet{2021A&A...646A.104H}, from which the ones we were able to detect have mean dispersions of $\overline{\sigma_{\varpi}} = 0.03$ mas and $\overline{\sigma_{\mu_{\alpha^*}}}, \overline{\sigma_{\mu_\delta}} = 0.11$ mas yr$^{-1}$, and the cluster that were not re-detected have $\overline{\sigma_{\varpi}} = 0.05$ mas and $\overline{\sigma_{\mu_{\alpha^*}}}, \overline{\sigma_{\mu_\delta}} = 0.18$ mas yr$^{-1}$, showing that the clusters we are able to recover are more compact.

Comparisons to pre-\textit{Gaia} cluster catalogues are more difficult because they do not allow for sufficiently good comparisons in proper motion space. These catalogues contained around $3\,000$ catalogued objects gathered from different data sources \citep{dias,kharchenko}. Most of their reported clusters identified with our findings were taken into account when crossmatching with the catalogue by \citet{tristan_catalogue} since they are based on the same data. For the clusters not found within \citet{tristan_catalogue}, we performed a $10$ arcmin positional crossmatch based on sky coordinates only. We flagged the coinciding candidates in our main Table~\ref{tab:oc_params} (see Sect.~\ref{subsec:new_ubc}); however, we did not explore further coincidences in any other dimension. 

\begin{figure}
\centering
\includegraphics[width = 1.\columnwidth]{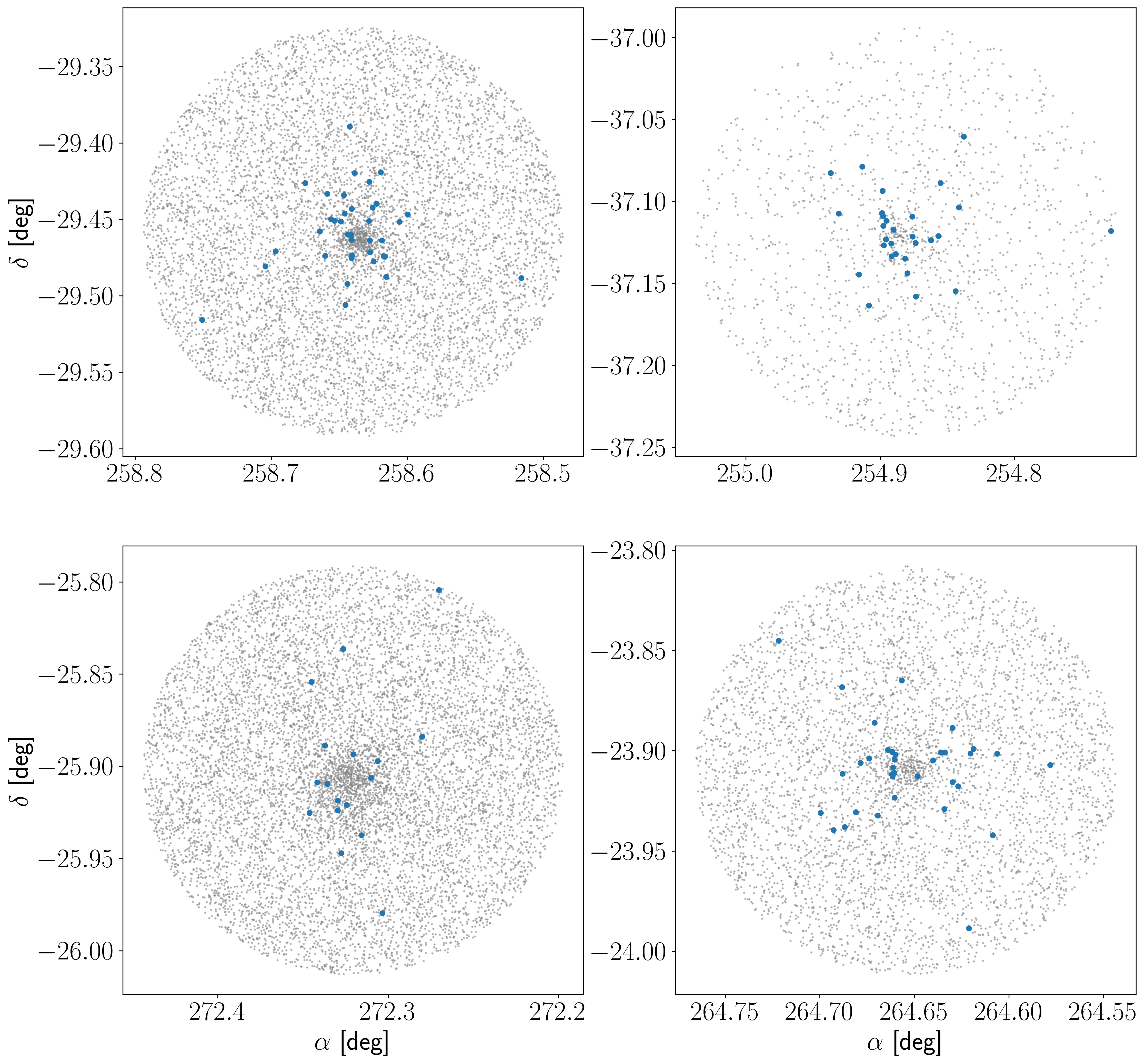}
\caption{Distribution in $\alpha$ and $\delta$ for GCs NGC~$6304$ (top left), NGC~$6256$ (top right), NGC~$6553$ (bottom left), and NGC~$6401$ (bottom right). The blue dots represent the stars found as overdensities with \texttt{OCfinder}, while grey dots are all stars in a cone search around the GC centre.}
\label{fig:globulars}
\end{figure}

Similarly to the OC catalogues, we crossmatched our findings to globular cluster (GC) catalogues. Recently, \citet{2021MNRAS.505.5978V} reported a catalogue of known GCs with mean astrometric parameters computed from \textit{Gaia} EDR3. There are $113$ GCs that can be found by \texttt{OCfinder} within our filters. We were able to find $94$ of them in the clustering step, which represents $83\%$ of the target catalogue, showing a re-detection efficiency similar to the OC case. Out of these, $84$ GCs were validated with the ANN in our photometric step. This shows that the addition of the simulated isochrones in the ANN training (see Sect~\ref{subsec:ann}), which are the only contribution for clusters with these old ages, improves the validation of cluster CMDs. There are four cases that escaped the crossmatch with a known GC due to differences in parallax and proper motions higher than our threshold. These cases are shown in Fig.~\ref{fig:globulars}, where we show all stars (in grey) around the centres of NGC~$6304$, NGC~$6256$, NGC~$6553$, and NGC~$6401$ as well as the cluster stars we were able to find around them (in blue). In these cases, we were able to find both the main host GC (already crossmatched) and a structure which is more dispersed than the catalogued GC, with differences of $3\sigma$ either in $\varpi$, $\mu_{\alpha^*}$, or $\mu_\delta$, and with a CMD compatible with being a very old object. Therefore, we consider these objects to be part of the same structure as the host GC, revealing the presence of extensive halos around these distant objects.

\subsection{New UBC clusters}
\label{subsec:new_ubc}

\begin{figure*}
\centering
\includegraphics[width = 1.\textwidth]{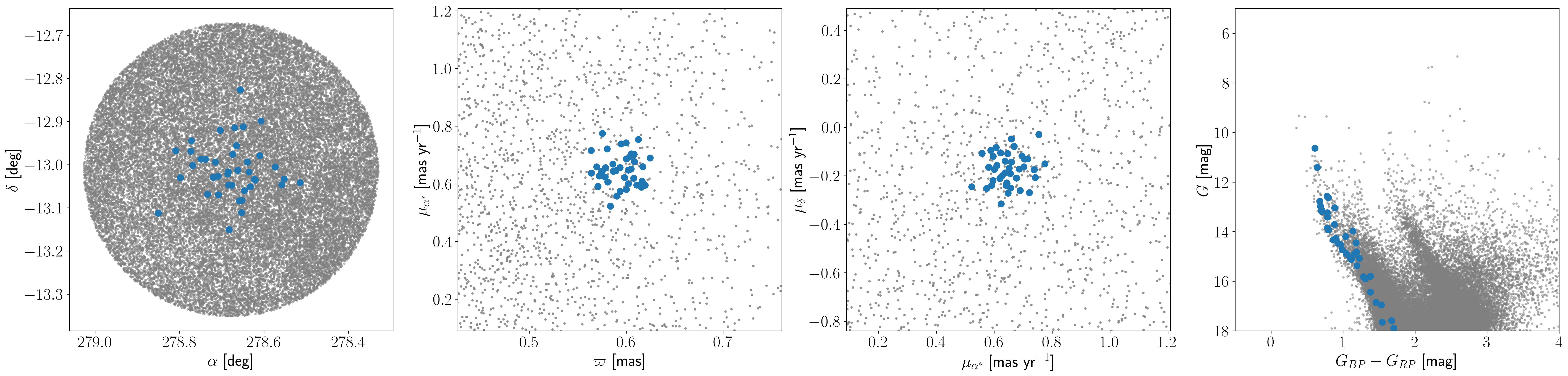}
\includegraphics[width = 1.\textwidth]{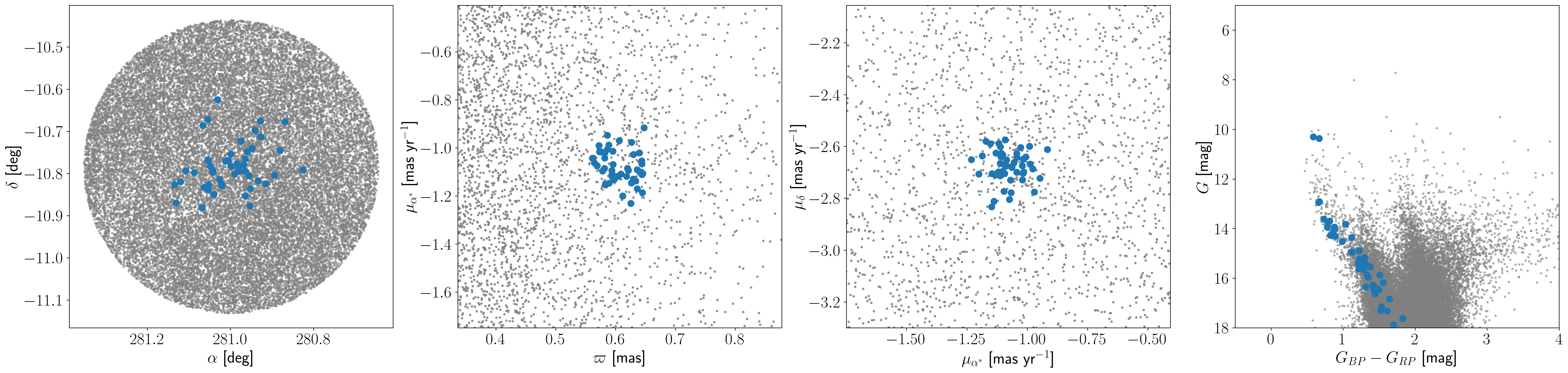}
\includegraphics[width = 1.\textwidth]{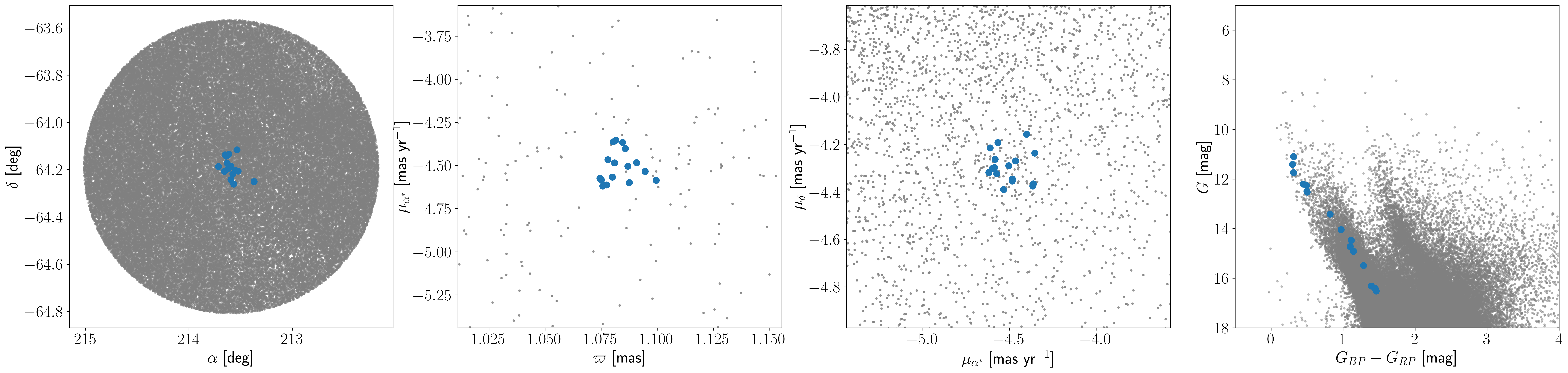}
\includegraphics[width = 1.\textwidth]{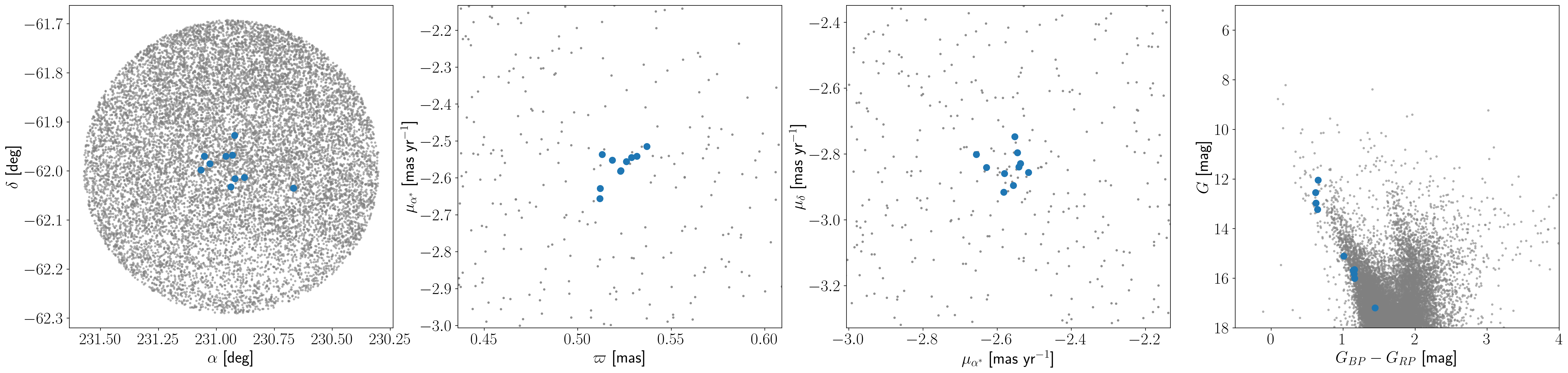}
\caption{Examples of the detected OCs for the different classes. The blue dots represent the detected member stars for each OC, while the grey dots are field stars queried in the \textit{Gaia} archive using a cone search within $10$ pc radius at the distance of the OC. From left to right, different panels represent : i) positional distribution in $\alpha$, $\delta$, ii) $\varpi$ versus $\mu_{\alpha^*}$ distribution, iii) proper motion diagram, and iv) the CMD. From top to bottom, different rows are for different OCs: i) UBC~$1023$ and UBC~$1029$ from class A, ii) UBC~$1592$ from class B, and ii) UBC~$1628$ from class C.}
\label{fig:cumuls}
\end{figure*}

After crossmatching our findings with known cluster catalogues, we were able to report $628$ new OC candidates, which are numbered from UBC~$1001$ in order to differentiate from the UBC clusters found in \textit{Gaia} DR2. These candidates were further divided into class A, class B, and class C based on a visual inspection of their distributions in $(\alpha,\delta,\varpi,\mu_{\alpha^*},\mu_\delta)$ and their CMDs, aided with the distribution of radial velocities when available. We classify $566$ OC candidates as class A ($90\%$ of the total), $26$ $(4\%)$ as class B, and $36$ $(6\%)$ as class C. Candidates in class A usually show a clustered distribution in the five astrometric dimensions and a clear sequence in the CMD. On the other hand, we generally classified candidates into class B if the main sequence formed by the candidate member stars in the CMD is truncated before $G = 18$ mag, and into class C if they also contain less than $15$ members (where we consider the validation to be less reliable due to small numbers). In Fig.~\ref{fig:cumuls} we show examples of class A (two first rows), class B (third row), and class C (fourth row) OC candidates, which give a visual idea of the features of OC candidates in  each of the classes.

A sample of the list of the new OC candidates, divided in their classes, can be found in Table~\ref{tab:oc_params}. It contains the mean astrometric parameters $(\alpha,\delta,l,b,\varpi,\mu_{\alpha^*},\mu_\delta)$ and their dispersions for each OC candidate together with radial velocities when available. It also contains the apparent angular radius $(\theta)$ of the OC candidate, computed as the quadratic sum of $\sigma_l$ and $\sigma_b$. A distance estimation computed from the CMD and mean parallax (see Sect.~\ref{subsubsec:ages}) is also provided, together with an estimation of age and line-of-sight extinction $(A_v)$. Finally, the number of stars considered as members, and member stars with available radial velocities, are also reported. We flagged the candidates that are positionally crossmatched to \citet{kharchenko} (see Sect.~\ref{subsec:comparison}). The full version of Table~\ref{tab:oc_params} can be found online at the CDS, together with Table~2 reporting the membership lists that resulted from our \texttt{OCfinder} method.

\begin{table*}
\begin{center}
\caption{Some examples of the OCs found in this paper.}
\centering
\scriptsize
\begin{adjustbox}{angle=90}
\begin{tabular}{lrrrrccrrccccc}
\hline
\hline
Name & $\alpha$ [deg] & $\delta$ [deg] & $l$ [deg] & $b$ [deg] & $\theta$ [deg] & $\varpi$ [mas] & $\mu_{\alpha^*}$ [mas yr$^{-1}$] & $\mu_{\delta}$ [mas yr$^{-1}$] & $V_{\text{rad}}$ [km s$^{-1}$] & $N$ ($N_{V_{\text{rad}}}$) & $\log{age}$ [dex] & $d$ [pc] & $A_v$ [mag] \\
\hline
\multicolumn{14}{c}{Class A} \\
\hline
UBC~$1001^{\text{a}}$&$267.36(0.10)$&$-29.23(0.04)$&$0.18(0.05)$&$-0.86(0.08)$&$0.10$&$0.37(0.01)$&$0.42(0.08)$&$-1.94(0.06)$&--(--)&29(0)&$7.46$&$2786$&$1.91$ \\
UBC~$1002$&$268.11(0.05)$&$-28.20(0.07)$&$1.40(0.06)$&$-0.90(0.05)$&$0.08$&$0.28(0.03)$&$0.01(0.06)$&$-1.96(0.06)$&-14.88(--)&71(1)&$8.34$&$3233$&$2.31$ \\
UBC~$1003$&$268.60(0.05)$&$-24.89(0.05)$&$4.48(0.05)$&$0.40(0.05)$&$0.07$&$0.39(0.02)$&$-0.58(0.09)$&$-3.11(0.06)$&--(--)&44(0)&$7.78$&$2809$&$2.66$ \\
UBC~$1004$&$270.56(0.07)$&$-25.07(0.06)$&$5.21(0.07)$&$-1.23(0.06)$&$0.09$&$0.38(0.04)$&$-0.90(0.11)$&$-2.40(0.10)$&-5.24(2.12)&418(2)&$8.50$&$2436$&$2.32$ \\
UBC~$1005^{\text{a}}$&$272.18(0.06)$&$-21.42(0.08)$&$9.12(0.08)$&$-0.75(0.06)$&$0.10$&$0.66(0.01)$&$-0.71(0.07)$&$-1.35(0.06)$&--(--)&31(0)&$7.37$&$1542$&$1.23$ \\
UBC~$1006$&$272.45(0.07)$&$-21.48(0.08)$&$9.19(0.07)$&$-0.99(0.07)$&$0.10$&$0.37(0.02)$&$0.07(0.08)$&$-1.06(0.07)$&10.21(0.59)&77(2)&$8.02$&$2987$&$2.28$ \\
UBC~$1007$&$270.80(0.05)$&$-19.35(0.05)$&$10.30(0.06)$&$1.40(0.05)$&$0.08$&$0.65(0.02)$&$0.05(0.07)$&$-1.08(0.09)$&--(--)&32(0)&$7.89$&$1602$&$1.93$ \\
UBC~$1008$&$273.34(0.06)$&$-19.03(0.08)$&$11.74(0.08)$&$-0.54(0.07)$&$0.10$&$0.34(0.02)$&$-0.84(0.06)$&$-1.82(0.06)$&--(--)&29(0)&$7.84$&$3182$&$1.28$ \\
UBC~$1009$&$273.50(0.06)$&$-19.05(0.07)$&$11.79(0.08)$&$-0.69(0.05)$&$0.09$&$0.33(0.02)$&$-0.36(0.09)$&$-1.19(0.09)$&--(--)&46(0)&$7.74$&$2753$&$1.27$ \\
UBC~$1010^{\text{a}}$&$274.50(0.06)$&$-18.54(0.07)$&$12.69(0.07)$&$-1.28(0.05)$&$0.09$&$0.37(0.02)$&$-0.01(0.05)$&$-0.50(0.08)$&--(--)&46(0)&$8.34$&$2634$&$1.08$ \\
UBC~$1011$&$274.10(0.06)$&$-18.11(0.07)$&$12.89(0.07)$&$-0.74(0.06)$&$0.09$&$0.34(0.04)$&$-1.11(0.10)$&$-2.64(0.09)$&9.39(8.93)&235(2)&$7.73$&$2804$&$2.74$ \\
UBC~$1012^{\text{a}}$&$273.03(0.05)$&$-15.98(0.04)$&$14.27(0.05)$&$1.18(0.04)$&$0.06$&$0.39(0.03)$&$0.36(0.09)$&$-0.69(0.09)$&--(--)&44(0)&$8.89$&$2871$&$2.39$ \\
UBC~$1013$&$273.13(0.11)$&$-15.33(0.11)$&$14.89(0.11)$&$1.41(0.10)$&$0.15$&$0.58(0.02)$&$-0.24(0.07)$&$-1.76(0.09)$&--(--)&50(0)&$7.04$&$1839$&$1.20$ \\
UBC~$1014$&$276.55(0.09)$&$-15.15(0.08)$&$16.61(0.09)$&$-1.42(0.07)$&$0.12$&$0.48(0.01)$&$-0.72(0.06)$&$-2.73(0.06)$&--(--)&28(0)&$6.99$&$2126$&$1.90$ \\
UBC~$1015$&$276.74(0.08)$&$-14.11(0.08)$&$17.62(0.08)$&$-1.09(0.08)$&$0.11$&$0.30(0.05)$&$-0.77(0.10)$&$-2.12(0.11)$&3.74(17.66)&565(4)&$8.40$&$3323$&$2.61$ \\
\multicolumn{14}{c}{$\vdots$} \\
\hline
\multicolumn{14}{c}{Class B} \\
\hline
UBC~$1567$&$269.01(0.06)$&$-29.51(0.04)$&$0.67(0.03)$&$-2.25(0.06)$&$0.06$&$0.38(0.02)$&$-0.10(0.09)$&$-0.47(0.07)$&--(--)&44(0)&$7.02$&$2929$&$1.94$ \\
UBC~$1568$&$283.34(0.03)$&$-9.31(0.03)$&$24.84(0.03)$&$-4.66(0.02)$&$0.04$&$0.43(0.01)$&$0.86(0.04)$&$-0.83(0.05)$&--(--)&10(0)&$7.54$&$2798$&$1.46$ \\
UBC~$1569$&$282.29(0.06)$&$-7.15(0.04)$&$26.31(0.05)$&$-2.75(0.06)$&$0.08$&$0.37(0.01)$&$-0.31(0.03)$&$-1.04(0.03)$&--(--)&11(0)&$8.24$&$2554$&$1.02$ \\
UBC~$1570$&$298.12(0.06)$&$32.54(0.05)$&$68.42(0.06)$&$2.74(0.04)$&$0.07$&$0.40(0.01)$&$-2.41(0.04)$&$-5.39(0.05)$&-2.46(--)&25(1)&$8.46$&$2761$&$1.30$ \\
UBC~$1571$&$314.26(0.15)$&$46.45(0.11)$&$86.91(0.10)$&$0.63(0.12)$&$0.15$&$0.18(0.04)$&$-2.95(0.10)$&$-3.62(0.13)$&-45.89(13.34)&233(6)&$8.91$&$5053$&$2.51$ \\
UBC~$1572$&$329.04(0.03)$&$53.98(0.03)$&$98.90(0.02)$&$-0.50(0.02)$&$0.03$&$0.45(0.01)$&$-2.24(0.03)$&$-2.40(0.05)$&--(--)&10(0)&$8.03$&$2320$&$1.55$ \\
UBC~$1573^{\text{a}}$&$343.45(0.08)$&$62.61(0.03)$&$109.86(0.04)$&$2.76(0.02)$&$0.05$&$1.18(0.02)$&$-1.32(0.11)$&$-2.70(0.10)$&--(--)&17(0)&$7.29$&$863$&$4.09$ \\
\multicolumn{14}{c}{$\vdots$} \\
\hline
\multicolumn{14}{c}{Class C} \\
\hline
UBC~$1593$&$267.03(0.05)$&$-29.01(0.04)$&$0.22(0.02)$&$-0.50(0.05)$&$0.06$&$0.58(0.01)$&$0.57(0.07)$&$-1.43(0.07)$&--(--)&15(0)&$8.04$&$1829$&$1.65$ \\
UBC~$1594$&$276.49(0.02)$&$-13.74(0.03)$&$17.83(0.03)$&$-0.71(0.02)$&$0.03$&$0.33(0.01)$&$-0.66(0.04)$&$-3.65(0.06)$&--(--)&13(0)&$8.21$&$3270$&$2.80$ \\
UBC~$1595$&$289.95(0.07)$&$-1.87(0.04)$&$34.50(0.03)$&$-7.16(0.08)$&$0.08$&$1.07(0.02)$&$-0.78(0.07)$&$-2.12(0.04)$&--(--)&11(0)&$8.10$&$1028$&$2.13$ \\
UBC~$1596$&$285.62(0.04)$&$11.89(0.04)$&$44.77(0.05)$&$2.98(0.03)$&$0.06$&$0.25(0.01)$&$-1.81(0.07)$&$-5.09(0.05)$&18.88(15.03)&13(2)&$8.55$&$4133$&$2.39$ \\
UBC~$1597$&$293.17(0.07)$&$16.81(0.07)$&$52.54(0.07)$&$-1.19(0.06)$&$0.09$&$0.34(0.01)$&$-2.28(0.04)$&$-5.59(0.05)$&--(--)&13(0)&$7.12$&$3225$&$3.04$ \\
UBC~$1598$&$313.90(0.03)$&$49.11(0.04)$&$88.78(0.04)$&$2.54(0.03)$&$0.04$&$0.36(0.02)$&$-2.71(0.04)$&$-4.54(0.03)$&-36.83(--)&10(1)&$9.04$&$3585$&$1.83$ \\
\multicolumn{14}{c}{$\vdots$} \\
\hline
\end{tabular}
\end{adjustbox}
\tablefoot{The different panels are for OCs classified in class A, class B, or class C, respectively. The parameters for each OC are the mean astrometric parameters and their standard deviations, computed from the $(N)$ member stars found in this work ($N_{V_{\text{rad}}}$ with radial velocity measurements). The apparent angular radius $(\theta)$ and an estimation for age, astro-photometric distance, and extinction are also included. The numeration starts from UBC~$1001$ to differentiate from UBC clusters found in \textit{Gaia} DR2. The full list can be found online at the CDS. $^{(\text{a})}$ For information regarding the positional coincidence with \citet{kharchenko}, see Sect.~\ref{subsec:comparison}}.
\label{tab:oc_params}
\end{center}
\end{table*}

\subsubsection{Characteristics of the new OC candidates}
\label{subsec:comments}

A small subset of the \textit{Gaia} EDR3 stars have radial velocity measurements from \textit{Gaia} DR2. We did not use these radial velocities in the clustering process; however, when available, they are useful to assess the reliability of the classification of the OC candidate. For the OC candidates in class A, $178$ of them have radial velocity measurements, of which $72$ are based on more than one star, and $25$ are based on more than two stars. From these $25$ OC candidates with radial velocities averaged over more than two stars, the median value of the radial velocity dispersions is $2.31$ km$\cdot$s$^{-1}$ with a median absolute dispersion of $2.03$ km$\cdot$s$^{-1}$, and $17$ of them have radial velocity dispersions of $3$ km$\cdot$s$^{-1}$ at most. For class B and class C candidates, only four and six clusters have a mean radial velocity available, respectively, with $1$ OC with more than two stars with radial velocity measurements in both cases. In these OCs, the radial velocity dispersions are $13.34$ and $15.03$ km$\cdot$s$^{-1}$, respectively. 

\begin{figure*}
\centering
\includegraphics[width = 1.\textwidth]{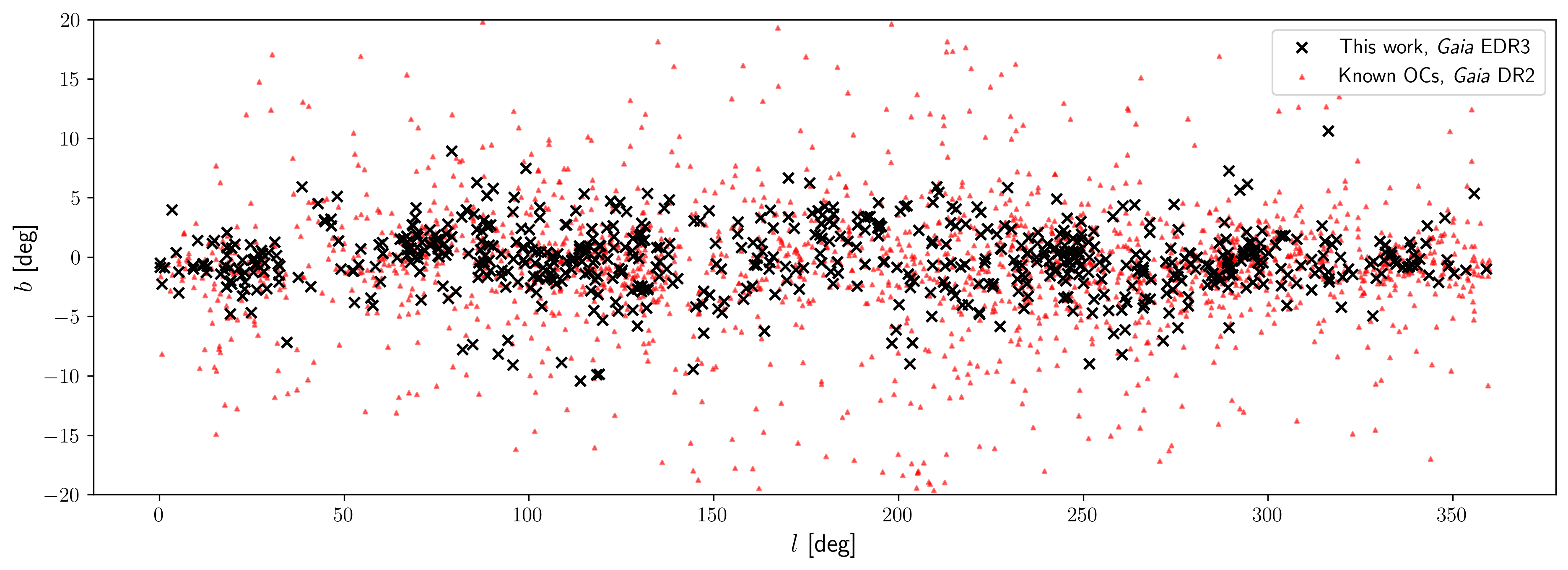}
\caption{Distribution of the OC population in $l$, $b$ Galactic coordinates. Red triangles represent the OCs known prior to this study, reported in \citet{2020A&A...640A...1C}. Black crosses represent the new OCs found in this work using \textit{Gaia} EDR3.}
\label{fig:lvsb}
\end{figure*}

Figure~\ref{fig:lvsb} shows the distribution of the new OC candidates in Galactic $l$ and $b$ coordinates. We see that the location of the new candidates matches that of the known OCs. Even if the search is performed up to $|b| \leq 20^\circ$, new OCs are preferentially located within the Galactic disc at low latitudes. The vast majority of the new OC candidates $(\sim 99\%)$ are located at $|b| \leq 10^\circ$ (with only UBC~$1186$ and UBC~$1530$ with $|b| > 10^\circ$, both belonging to class A), and $\sim 93\%$ of them within $|b| \leq 5^\circ$. We are also able to confirm some structures seen in previous studies, such as the lack of OCs in a region near $l \sim 140^\circ$ previously dubbed the Gulf of Camelopardalis \citep{coin_clusters,acastro2}.

\begin{figure}
\centering
\includegraphics[width = 1.\columnwidth]{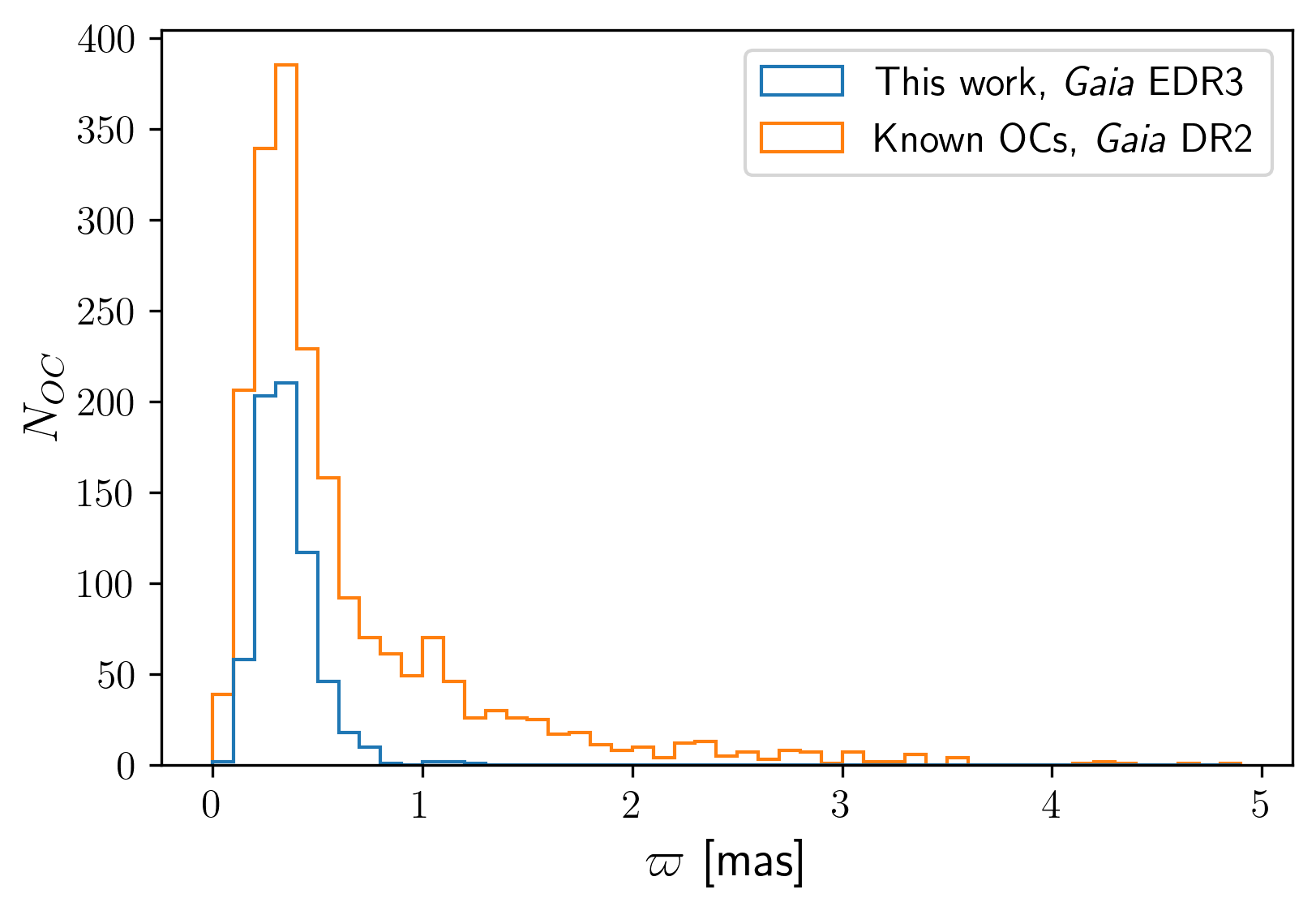}
\caption{Histogram of parallaxes for the OC population. The orange line shows the known population in \citet{2020A&A...640A...1C}, while the blue line shows the new findings in this study.}
\label{fig:parallax_hist}
\end{figure}

The fact that our new detections are generally at larger distances can be seen in Fig.~\ref{fig:parallax_hist}, where we show a histogram of the OC mean parallaxes for both the known population \citep{2020A&A...640A...1C} and new OC candidates. We see that the parallax distribution of the new OCs is positively skewed with respect to the distribution of known OC, meaning that the mode is slightly towards smaller mean parallaxes. Also, the drop in the distribution towards larger parallaxes is steeper in the new OCs' distribution, showing the increasing difficulty in finding new nearby ones. In fact, only four OC candidates $(0.60\%)$ are closer than $1$ kpc (see Sect.~\ref{subsec:nearby}), and $75$ $(11.3\%)$ are located within $1$ and $2$ kpc, probably due to a better completeness of previous surveys in these regions among other methodological effects. The relative parallax errors $(\sigma_\varpi/\varpi)$ for our new OCs range from $0.003$ to $0.05$ in the case of class A OCs and from $0.002$ to $0.02$ for class B and class C.  

\begin{figure}
\centering
\includegraphics[width = 1.\columnwidth]{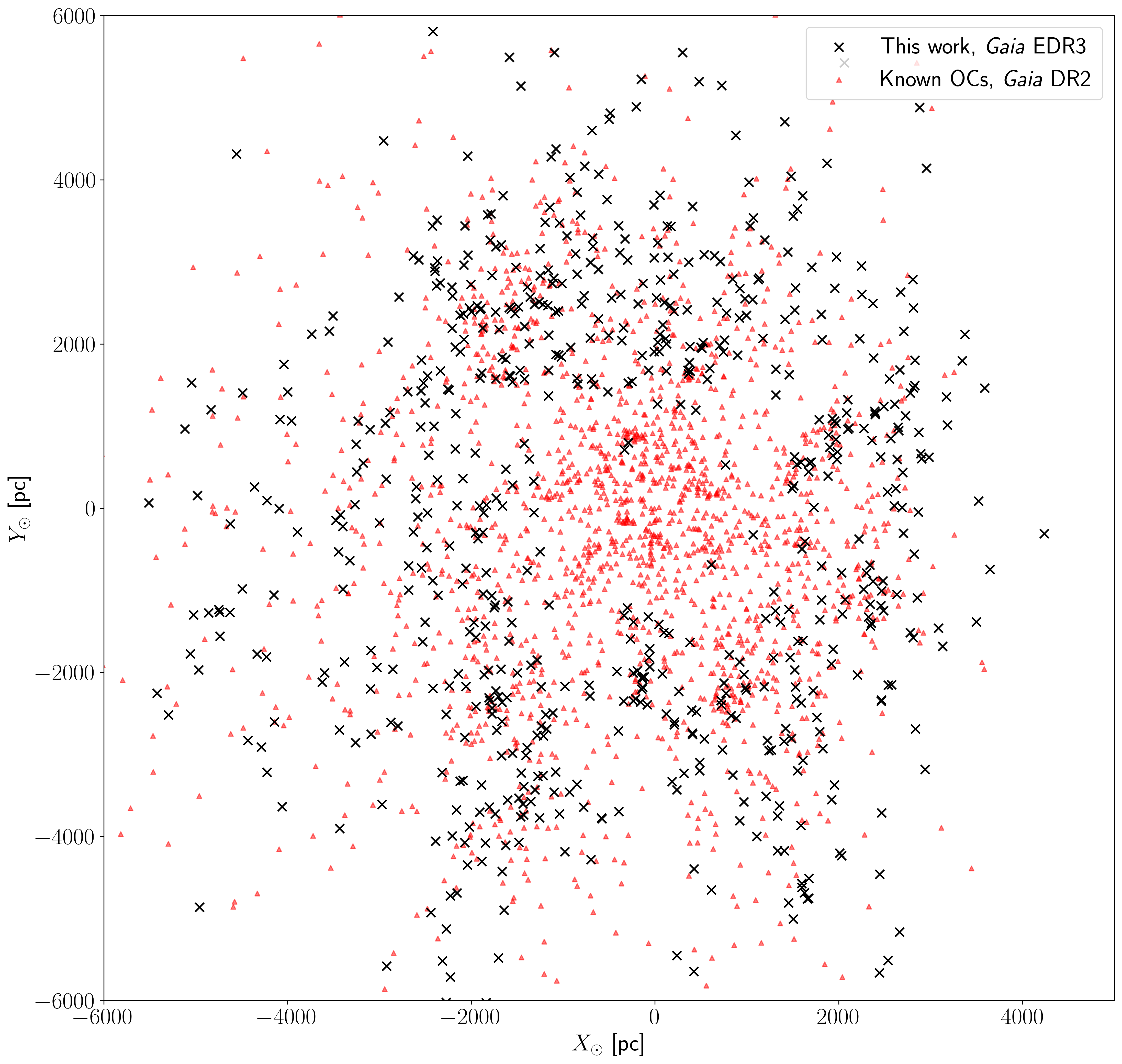}
\caption{Distribution in $X_{\odot}$ and $Y_{\odot}$ heliocentric coordinates. Symbols are the same as in Fig.~\ref{fig:lvsb}.}
\label{fig:x_y_solar}
\end{figure}

The heliocentric distances for the new OC candidates range from $860$ pc to $9.6$ kpc, computed from the distance modulus (see Sect.~\ref{subsubsec:ages}). In Fig.~\ref{fig:x_y_solar} we show the distribution in the $X_{\odot}$ and $Y_{\odot}$ coordinates, where we see that very few new OC candidates are detected within $1.5$ kpc. This may be due to the combination of the following: i) the approach adopted in \texttt{OCfinder}, where we are limited to the most compact object in the search region (see Sect.~\ref{subsec:clustering}), and those are more likely to be already known at these close distances; and ii) we expanded the search to $G = 18$ mag, which naturally pushes the search to farther distances \citep[see][to see the performance of \texttt{OCfinder} in terms of completeness for nearby objects]{2021A&A...645L...2A}, together with the above consideration of a better completeness of the nearby population.

\begin{figure}
\centering
\includegraphics[width = 1.\columnwidth]{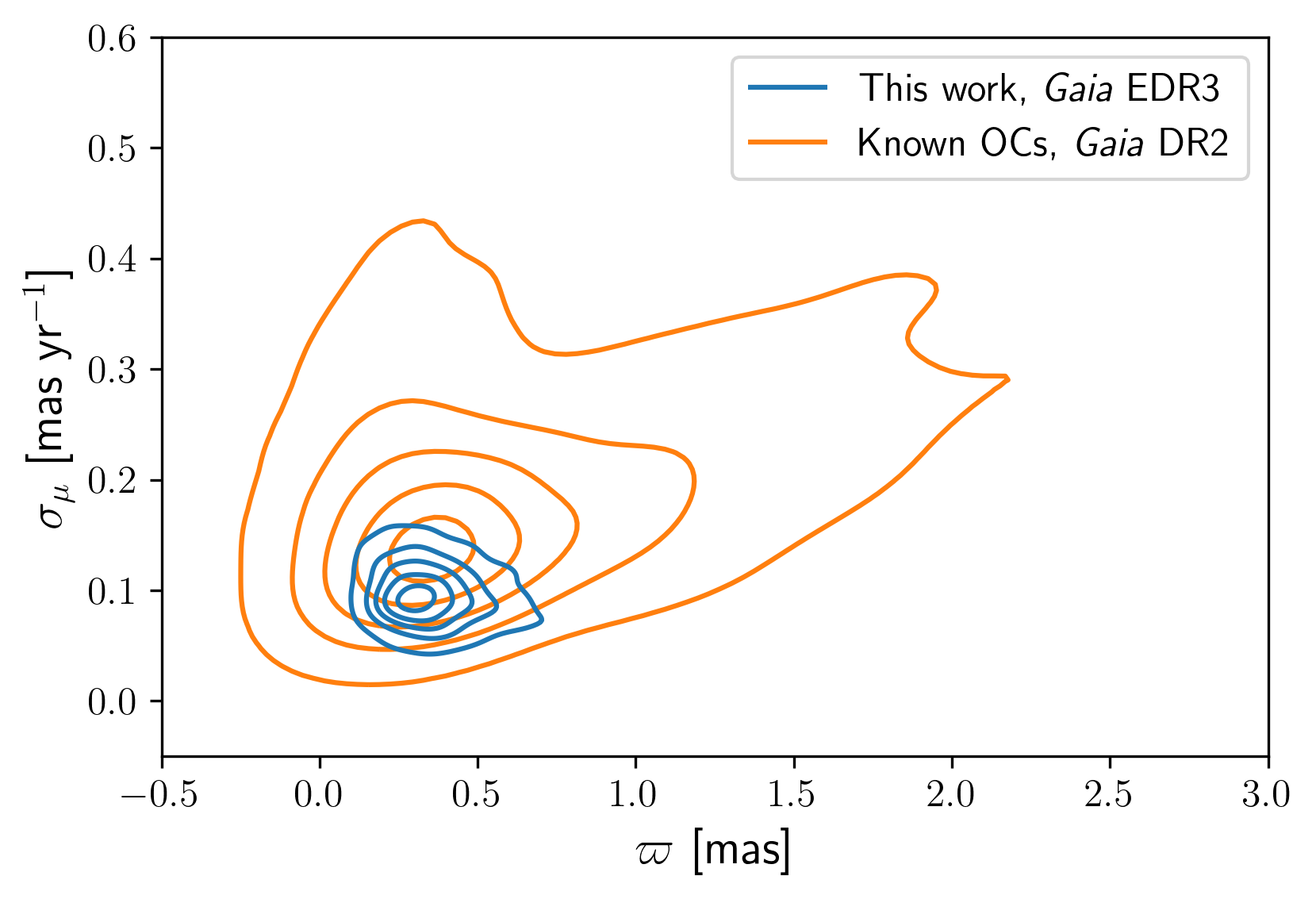}
\caption{Total proper motion dispersion $\left(\sqrt{\sigma_{\mu_{\alpha^*}}^2 + \sigma_{\mu_\delta}^2}\right)$ as a function of parallax. The orange lines show the OC density contours for the OC characterised with \textit{Gaia} DR2 \citep{2020A&A...640A...1C}. The blue lines show the same OC density contours for the OCs detected in this study using \textit{Gaia} EDR3. The density contours are at the $10$, $30$, $50$, $70$, and $90\%$ levels.}
\label{fig:pm_parallax_density}
\end{figure}

The effect of the improvements in the \textit{Gaia} EDR3 data is also seen in Fig.~\ref{fig:pm_parallax_density}. There, we show a contour plot of the total proper motion dispersion as a function of the mean parallax, mimicking Fig.~1 in \citet{2020A&A...633A..99C}. The densest part of the distribution, where most of the clusters are, is moved towards smaller $\sigma_\mu$ showing the huge improvement in the proper motion determinations (the effect is smaller in the parallax). This results in OCs being more compact (in proper motion and parallax), and thus it is easier to detect them as overdensities with respect to \textit{Gaia} DR2.

\subsubsection{Ages, distances, and line-of-sight extinctions}
\label{subsubsec:ages}

\citet{2020A&A...640A...1C} trained an ANN on a set of well-characterised OCs with reliable estimations for ages, distances, and line-of-sight extinctions to estimate these parameters for almost the whole OC population characterised with \textit{Gaia} DR2 data. We fine-tuned this ANN to estimate these astrophysical parameters for the newly discovered OCs with \textit{Gaia} EDR3 data and we include them in our Table~\ref{tab:oc_params}. The ANN takes the CMD of the OC member stars into account, together with the mean parallax plus two other quantities derived from the CMD to aid in the estimation \citep[see Sect.~3.1 from][for details]{2020A&A...640A...1C}. For each OC, the ANN estimates its age, absorption ($A_v$), and the distance modulus, and from this we were able to estimate the distance. The authors compared the values from a set of reference clusters with their estimated values to account for their uncertainties. They report that the uncertainties on the determination of the $\log{age}$ range from $0.15$ to $0.25$ dex for young OCs ($\leq 8.5$ dex), and from $0.1$ to $0.2$ dex for old OCs. In the case of extinction and distance modulus, the reported typical uncertainties range from $0.1$ to $0.2$ mag for $A_v$, and from $0.1$ to $0.2$ mag in the distance modulus which corresponds to a $5\%$ to $10\%$ distance uncertainty. For further details, readers can refer to Sect.~3.4 from \citet{2020A&A...640A...1C}.

\begin{figure}
\centering
\includegraphics[width = 1.\columnwidth]{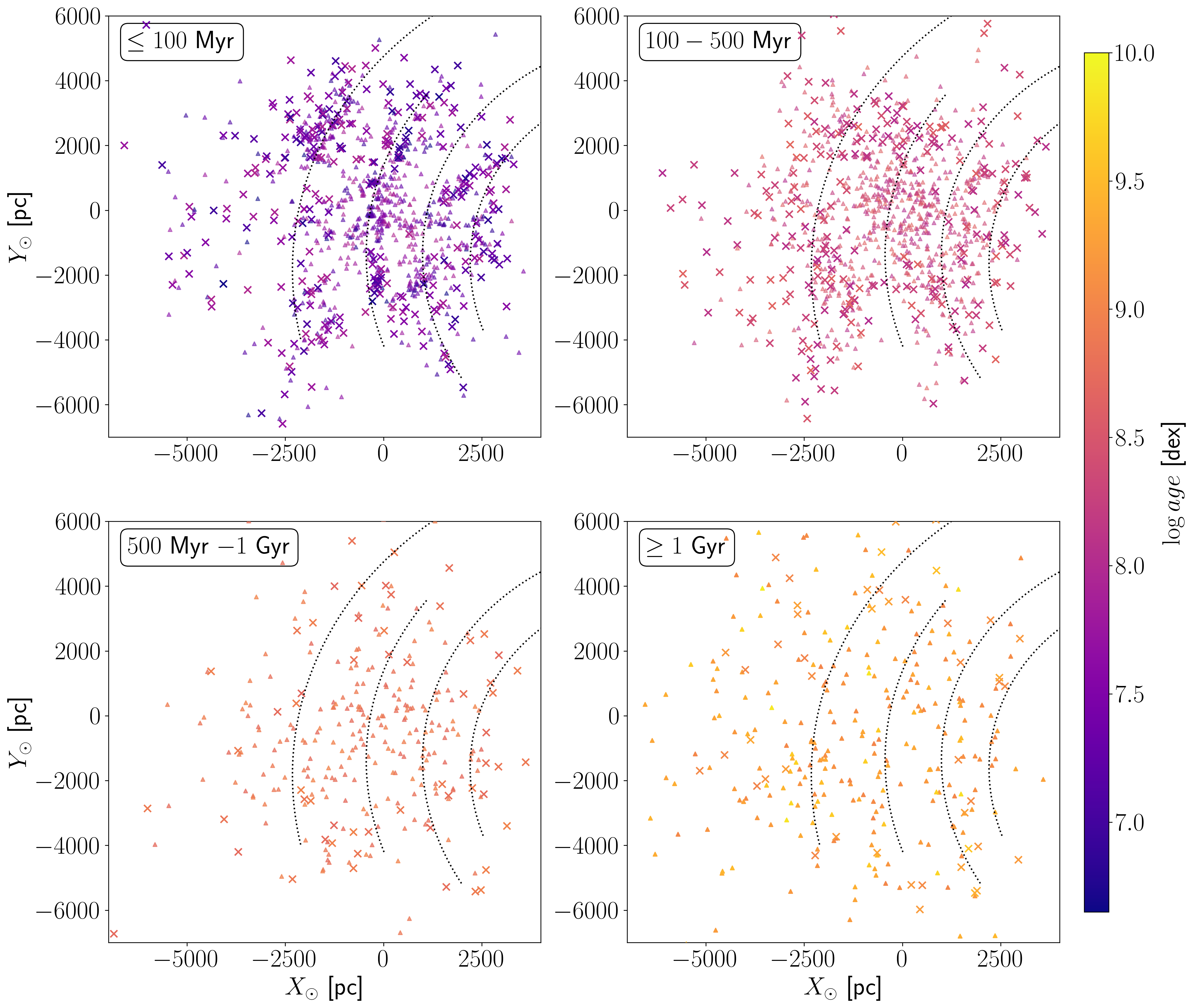}
\caption{Distribution of the new (crosses) and known \citep[triangles,][]{2020A&A...640A...1C} OCs in the $X_{\odot}$ versus $Y_\odot$ coordinates for different age bins: $0-100$ Myr (top left), $100-500$ Myr (top right), $500$ Myr $-$ $1$ Gyr (bottom left), and more than $1$ Gyr (bottom right). The dotted lines show the spiral arms as described by \citet{2021A&A...652A.162C}.}
\label{fig:age_bins}
\end{figure}

In Fig.~\ref{fig:age_bins} we plotted the distribution of the new OCs (crosses), together with the previously known population (triangles), in the Galactic disc for different age intervals: i) younger than $100$ Myr counting with $276$ and $703$ new and previously known OCs, respectively; ii) from $100$ to $500$ Myr, with $248$ new OCs and $675$ previously known OCs; iii) $500$ Myr to $1$ Gyr, $58$ new OCs and $229$ previously known OCs; and iv) older than $1$ Gyr, with $46$ new OCs and $260$ previously known OCs. These OCs are also colour-coded by their age. In the younger age bin ($\leq 100$ Myr), we find clear overdensities of OCs, which are following the different spiral arms \citep[as fitted by][]{2021A&A...652A.162C}. In the following age intervals, the distribution of the OCs is more dispersed, not showing significant overdensities, as expected. 

\begin{figure*}
\centering
\includegraphics[width = 1.\textwidth]{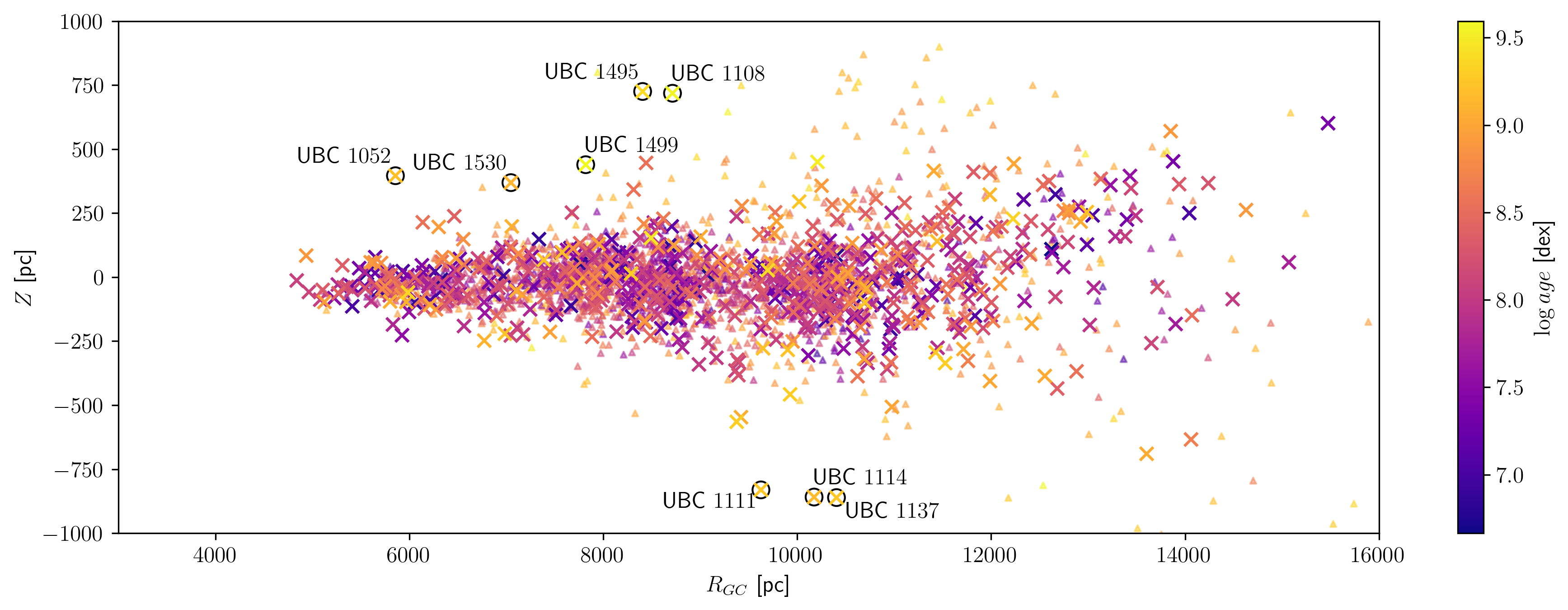}
\caption{Distribution of the newly found (crosses) and known (triangles) OCs in the $R_{GC}$ and $Z$ coordinates, colour-coded by age. The black circles represent the selected OC shown in Fig.~\ref{fig:highz}.}
\label{fig:r_z}
\end{figure*}

\begin{figure*}
\centering
\includegraphics[width = 1.\textwidth]{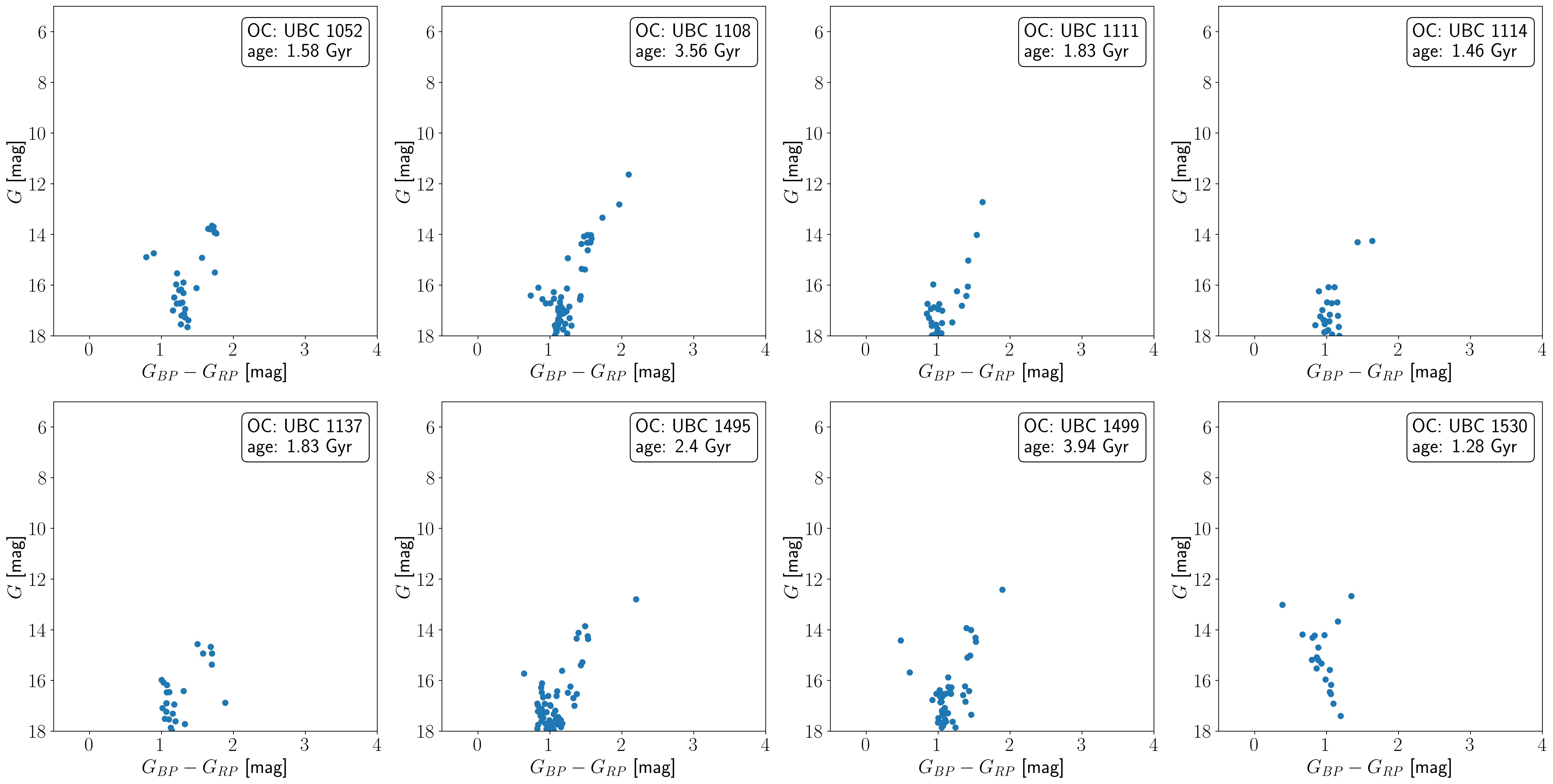}
\caption{CMDs of the high $|Z|$ OCs, selected in Fig.~\ref{fig:r_z}.}
\label{fig:highz}
\end{figure*}

\begin{figure}
\centering
\includegraphics[width = 1.\columnwidth]{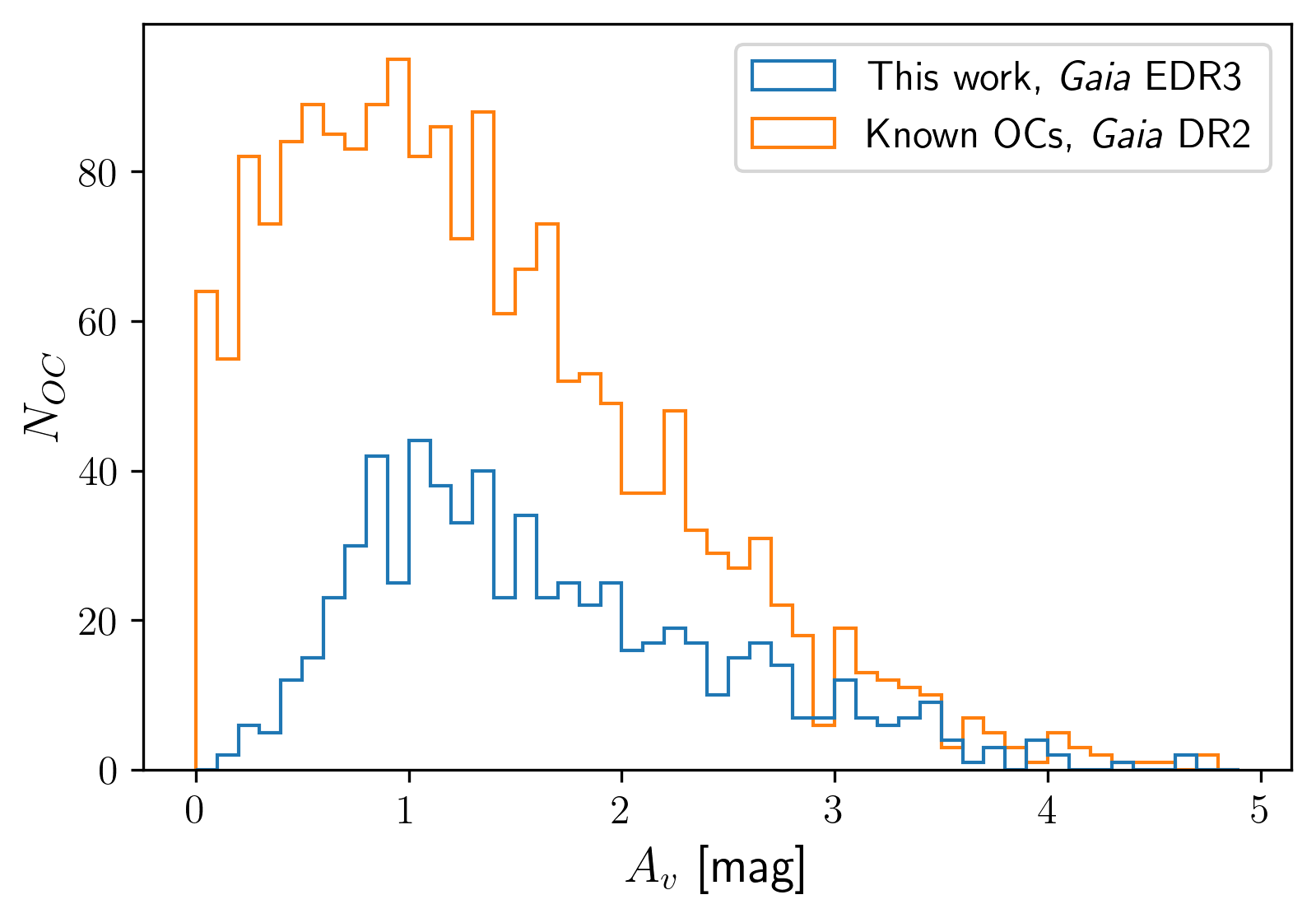}
\caption{$A_v$ histogram for the known (orange) and the new OC (blue) population.}
\label{fig:avhist}
\end{figure}

The consistency of the age estimations with the previously known OC population is also shown in Fig.~\ref{fig:r_z}, where we show the distribution of the new OCs in the galactocentric radius $(R_{GC})$ and altitude above the Galactic plane $(Z)$ coordinates. We did not find young open clusters at high $|Z|$ in the inner disc, but old OCs. The black circles highlight the OCs identified as old and with high $|Z|$, and specific plots for those are shown in show Fig.~\ref{fig:highz}. These are interesting OCs, all of them are older than $1$ Gyr and up to $4$ Gyr (which is the oldest OC in our findings), and follow-up studies will be needed to explore their nature. At large $R_{GC}$, we are also able to see the flare of the Galactic disc already seen with the OC population after \textit{Gaia} DR2 \citet{2020A&A...640A...1C}. The extinction values range from $0.14$ to $4.65$ mag, with a distribution that is shown in Fig.~\ref{fig:avhist}.

\subsubsection{Comments on UBC~$1061$}
\label{subsec:disrupt}

\begin{figure}
\centering
\includegraphics[width = 1.\columnwidth]{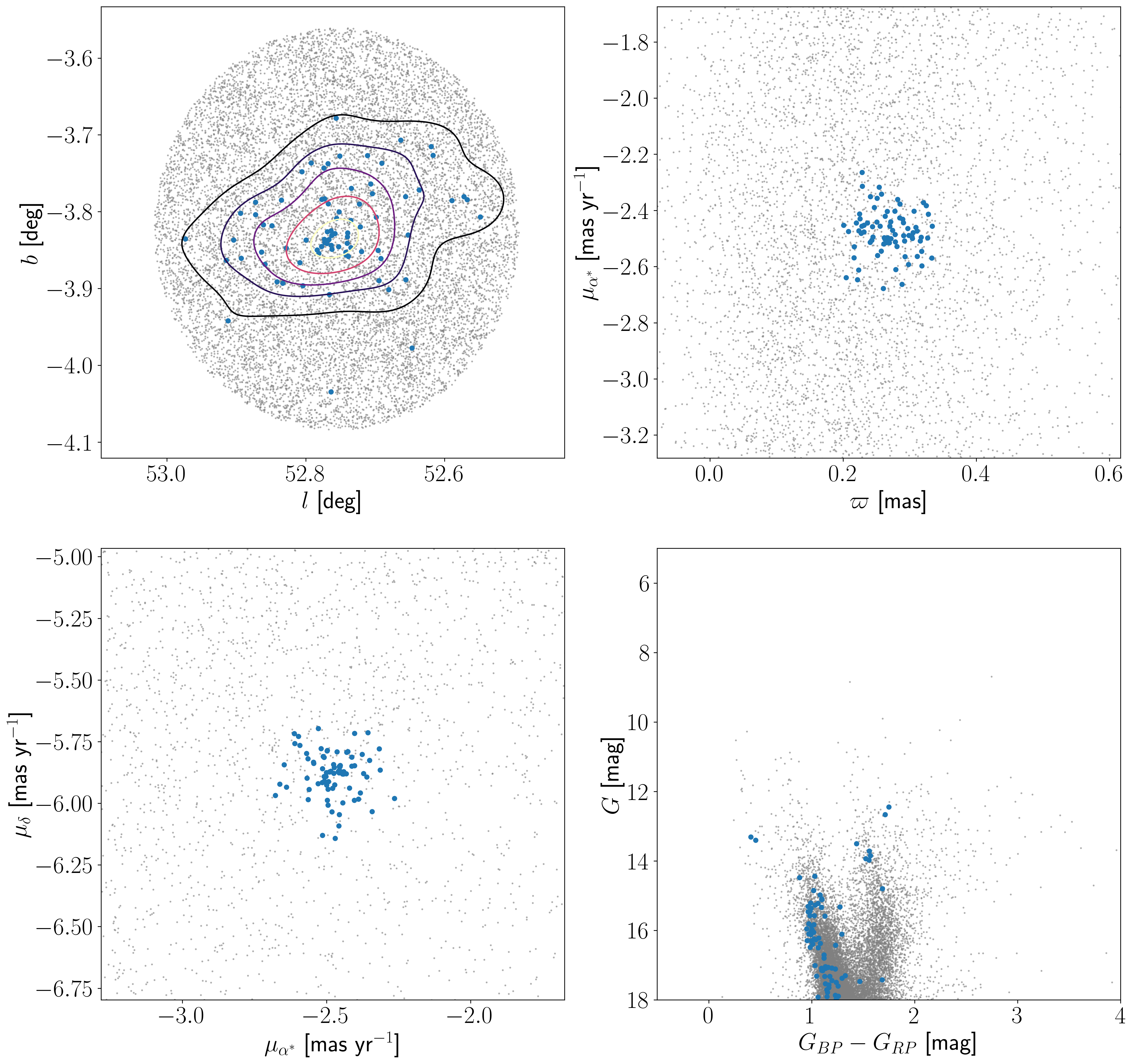}
\caption{Diagrams for UBC~$1061$. The blue dots are the stars selected as members with our \texttt{OCfinder} method, while the grey dots are field stars around $15$ pc from the cluster centre. The diagrams correspond to the sky distribution, with density contour plots (top left), $\varpi$ versus $\mu_{\alpha^*}$ (top right), proper motion diagram (bottom left), and the CMD (bottom right).}
\label{fig:ubc3328}
\end{figure}

In Paper III, we reported the discovery of UBC~$274$, an old OC with an extended profile due to disruption by tidal forces. Here, we were also able to detect stellar groups that are probably undergoing disruption processes. This is the case for UBC~$1061$, which is a new OC found at $l = 52.75^\circ$ and $b = -3.82^\circ$. It is located at a galactocentric distance of $R_{GC} = 6.77$ kpc, associated with the Sagittarius arm \citep{2014ApJ...783..130R,2021A&A...652A.162C}, and at $Z = -247.95$ pc. We estimate the age of UBC~$1061$ to be $1.3$ Gyr, with an extinction value of $A_v = 0.74$ mag. Figure~\ref{fig:ubc3328} shows the distribution of its member stars in five astrometrical dimensions, in addition to its CMD where two blue straggler star candidates can be seen. While in parallax and proper motion UBC~$1061$ shows a clustered structure, it presents an elongation in the sky position diagram along the $l$ coordinate, probably because it is undergoing disruption processes by the tidal forces \citep[see][for more examples]{2019A&A...621L...2R,2020A&A...639A..55P}. Such an old OC is rare in current OC catalogues, fewer than $20\%$ of the reported OCs are older than $1$ Gyr, and this is particularly the case for an inner disc cluster where young OCs are more frequent \citep{2020A&A...640A...1C}. None of the $79$ identified member stars of UBC~$1061$ have radial velocities in \textit{Gaia} EDR3, and its red clump stars (at $G \sim 14$ mag) are at the faint limit for the $G_{\rm RVS}$ sample. Therefore, UBC~$1061$ is an interesting case to follow up with on-ground spectroscopic surveys.

\subsubsection{Candidates within $1$ kpc}
\label{subsec:nearby}

We found four OC candidates within $1$ kpc from the Sun, where the OC census was assumed to be complete before \textit{Gaia}. These OC candidates are UBC~$1187$, UBC~$1573,$ and UBC~$1592$, which are located at $861$, $862,$ and $977$ pc, respectively, and all of them have angular sizes $\leq 0.10^\circ$. They are young objects, with none of them showing red clump stars in their CMDs, and they are poorly populated, with $\sim 15$ stars per cluster. This confirms the assumption that our methodology \texttt{OCfinder} is better suited to find objects with small angular sizes, and thus farther objects. However, since nearby objects are still detected, this opens the possibility of performing dedicated searches in the Solar neighbourhood which will be enhanced with the future \textit{Gaia} DR3 thanks to the $\sim 33\cdot10^6$ stars with radial velocity measurements\footnote{\url{https://www.cosmos.esa.int/web/gaia/dr3}}.

\section{Conclusions}
\label{sec:conclusions}

We report the detection of $628$ new OCs within de Galactic disc, as the result of the adaptation and application of the \texttt{OCfinder} method to \textit{Gaia} EDR3. For all of the new OCs, we report mean astrometric values $(\alpha,\delta,l,b,\parallax,\mu_{\alpha^*},\mu_\delta)$ and radial velocity when available. We also estimate their ages, distances, and line-of-sight extinctions using an ANN trained on well-characterised OCs that was successfully applied to the OCs known in \textit{Gaia} DR2. We divided the new OCs into class A, class B, and class C depending on the reliability of the candidate by inspecting the distribution of their member stars in all the available dimensions.

We find that the new OCs are located, in general, at farther distances than the clusters known from \textit{Gaia} DR2. In fact, we were only able to detect three new objects closer than $1$ kpc from the Sun. This is thanks to the improvements in parallax and proper motion precisions of \textit{Gaia} EDR3, with respect to DR2 which allowed us to find more clustered objects (and thus we were able to search farther), and better knowledge of the OC population at close distances.

The estimation of astrophysical parameters also adds reliability to the OCs found with \texttt{OCfinder}. We see that young OCs follow the Galactic spiral arms, and they disperse on the Galactic disc as we explore older ages. Also, we find most OCs to be located at low $|Z|$ in the inner disc, with the exception of some old ($> 1$ Gyr) newly found OCs. In the outer disc, we are able to see the flaring of the Galactic disc with young OCs.

The use of a big data environment in the \texttt{OCfinder} methodology is key for a successful search of OCs. So far, $1\,274$ OCs have been discovered using the \texttt{OCfinder} method, which represents almost $50\%$ of the currently known OCs population. We have shown that improvements in the OCs census, both in terms of new detections or characterisation of known OCs, need to be aided by machine-learning methods to extract knowledge of the huge volume of high-quality data that is provided by \textit{Gaia} EDR3. Future \textit{Gaia} releases, as well as future photometric and spectroscopic Galactic surveys, will only increase this huge volume of high-quality data and thus the need for machine-learning methods.

\begin{acknowledgements}

This work has made use of results from the European Space Agency (ESA)
space mission {\it Gaia}, the data from which were processed by the {\it Gaia
Data Processing and Analysis Consortium} (DPAC).  Funding for the DPAC
has been provided by national institutions, in particular the
institutions participating in the {\it Gaia} Multilateral Agreement. The
{\it Gaia} mission website is \url{http: //www.cosmos.esa.int/gaia}. The
authors are current or past members of the ESA {\it Gaia} mission team and
of the {\it Gaia} DPAC.

This work was (partially) funded by the Spanish MICIN/AEI/10.13039/501100011033 and by "ERDF A way of making Europe" by the “European Union” through grant RTI2018-095076-B-C21, and the Institute of Cosmos Sciences University of Barcelona (ICCUB, Unidad de Excelencia ’Mar\'{\i}a de Maeztu’) through grant CEX2019-000918-M. 
ACG acknowledges Spanish Ministry FPI fellowship n. BES-2016-078499.

This work has been partially supported by the Spanish Government (PID2019-107255GB), by Generalitat de Catalunya (contract 2014-SGR-1051).

This research has made use of the VizieR catalogue access tool, CDS,
Strasbourg, France. The original description of the VizieR service was
published in A$\&$AS 143, 23.

This research has made extensive use of the TOPCAT software \citep{topcat}.
\end{acknowledgements}

\bibliographystyle{aa} 
\bibliography{bibliography}

\begin{appendix}
\section{Fitting magnitude uncertainties in \textit{Gaia}~EDR3}
\label{app:photometry_error}

The \textit{Gaia}~EDR3 photometric error model used in this work is based on some relatively simple analytical functions derived fitting magnitude uncertainties for one million random sources in the \textit{Gaia}~EDR3 catalogue, covering all magnitude ranges. This error model was used to produce mock \textit{Gaia} CMDs of the synthetic clusters used in the training of the ANN, described in Sect.~\ref{subsec:ann}.

\begin{figure}[htb!]
        \begin{center}
       \includegraphics[width=1.\columnwidth]{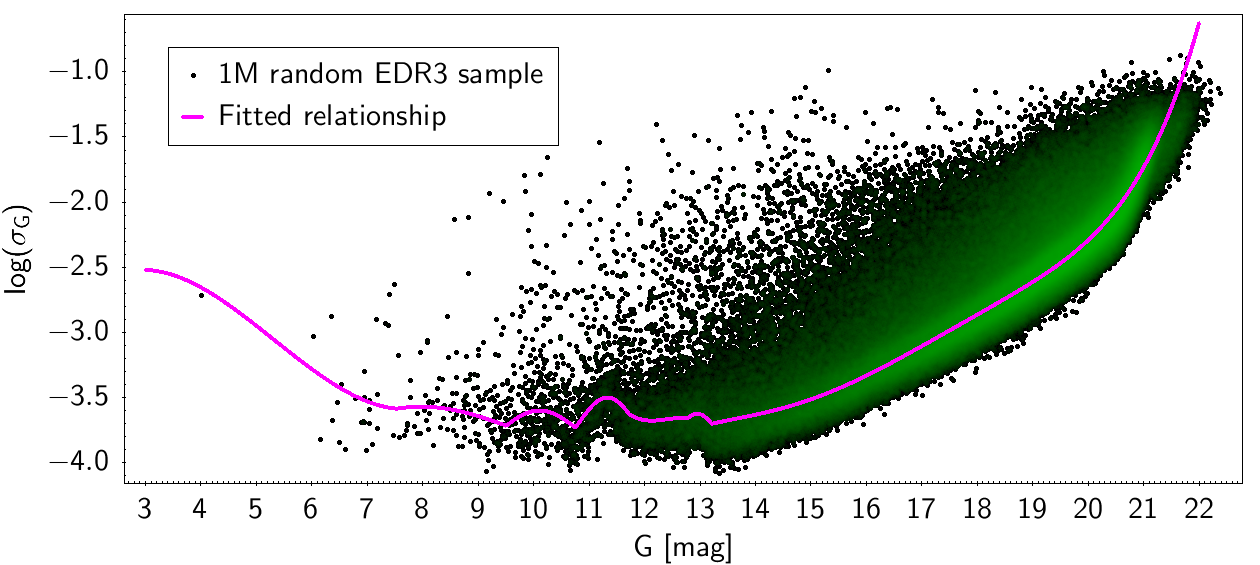}
       \includegraphics[width=1.\columnwidth]{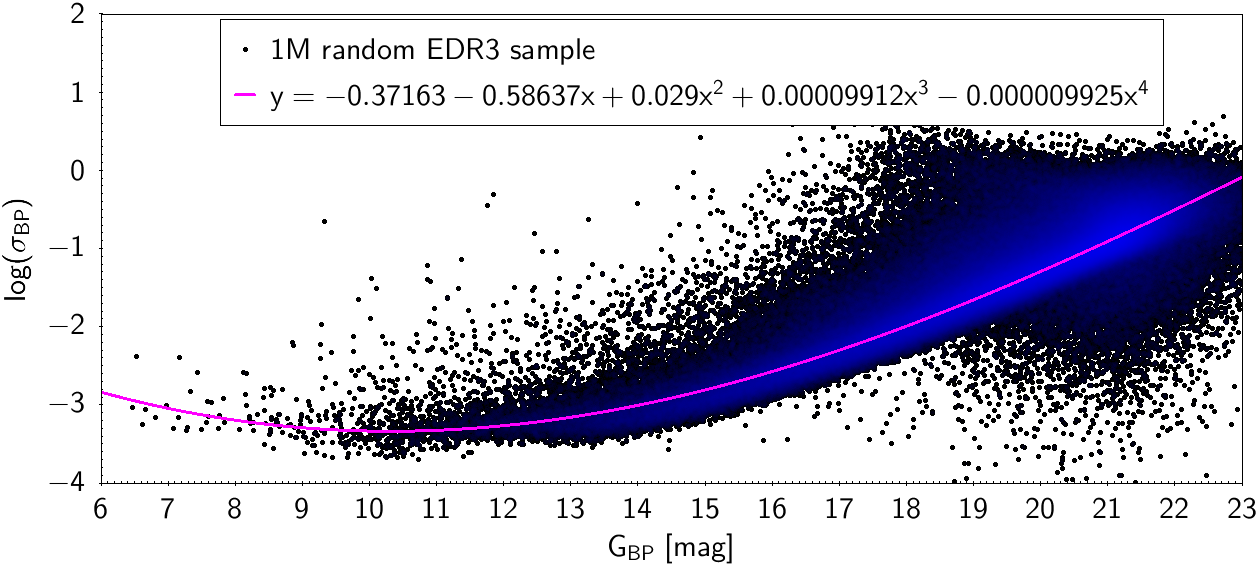}
       \includegraphics[width=1.\columnwidth]{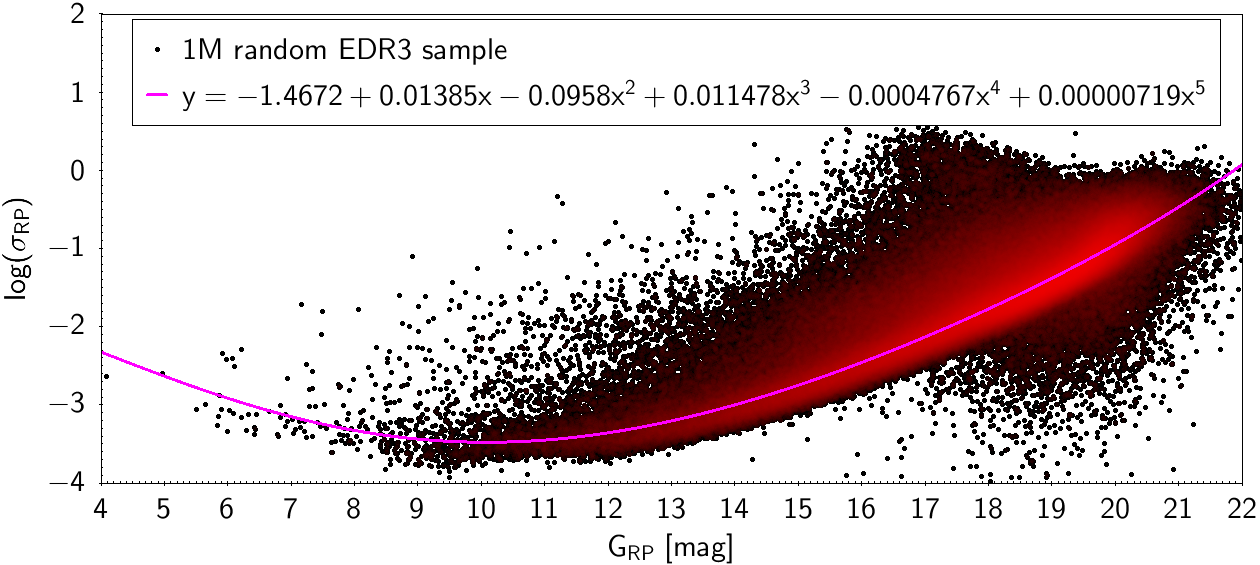}
        \caption{\textit{Gaia}~EDR3 magnitude uncertainties for $G$ (top), $G_{\rm BP}$ (centre), and $G_{\rm RP}$ (bottom) and the fitted laws included in Tables~\ref{tab:sigmafit}.
        \label{fig:sigma} 
        }
        \end{center}
\end{figure}

In order to fit the uncertainties, we assumed a polynomial relationship between the logarithm of the \textit{Gaia}~EDR3 magnitude uncertainties, $\log(\sigma_{G_{\rm XP}})$, and their magnitudes, following the expression
\begin{equation}
\log(\sigma_{G_{\rm XP}}) = \sum A_k \cdot (G_{\rm XP})^k,
\label{eq:polinomi}
\end{equation}
with $G_{\rm XP}$ being either $G$, $G_{\rm BP}$, or $G_{\rm RP}$ in every case. The resulting coefficients derived are shown in Table~\ref{tab:sigmafit}.

For the $G$ case, in some magnitude ranges the uncertainties in \textit{Gaia}~EDR3 show some complicated features due to calibration issues \citep[see][]{2021A&A...649A...3R}, which cannot be fitted with a simple polynomial.
A Gaussian function (Eq.~\ref{eq:gaussian}) was used instead of the polynomial (Eq.~\ref{eq:polinomi}) in these ranges:
\begin{equation}
\log(\sigma_G) = a+\exp\left(\frac{-(G-b)^2}{2c^2}\right). \\
\label{eq:gaussian}
\end{equation}

The actual fit used the following equation obtained taking the natural logarithm of Eq.~\ref{eq:gaussian}:
\begin{equation}
\ln[\log(\sigma_G)+4] = P \cdot G^2 + Q \cdot G + R,
\label{eq:gaussianpolyn}
\end{equation}
where we added 4 to $\log(\sigma_G)$ in order to avoid negative values when deriving its neperian logarithm. In order to retrieve the coefficients in Eq.~\ref{eq:gaussian} from the fitted coefficients (Table~\ref{tab:sigmafit}), the following transformation can be done from the values of $P$, $Q$, and $R$:
\begin{eqnarray}
a&=&e^{\left(R+\frac{Q}{2}\right)},\\
b&=&\frac{-Q}{2P},\\
c&=&\sqrt{\frac{-1}{2P}}.
\end{eqnarray}


\begin{table*}
\begin{center}
\caption{
Coefficients of the polynomial (Eq.~\ref{eq:polinomi}, first rows in the table with parameters $A_k$) and Gaussian (Eq.~\ref{eq:gaussian}, second part of the table with parameters $P$, $Q$, and $R$) equations for each magnitude range.}
\small
\begin{tabular}{cccccccc
}
\hline
Fitted uncertainty & Magnitude range & $A_0$ & $A_1$ & $A_2$ & $A_3$ & $A_4$ & $A_5$
 \\
\hline
$\log(\sigma_{\rm BP})$& $6<G_{\rm BP}<21$  &      
-0.37163 &-0.58637 & 0.029 &0.00009912 & -0.000009925 & -\\
$\log(\sigma_{\rm RP})$&$6<G_{\rm RP}<21$         & 
-1.4672 &0.01385 & -0.0958 &0.011478 & -0.0004767 &0.00000719\\
\hline          
Fitted uncertainty & Magnitude range & $A_0$ & $A_1$ & $A_2$ & $A_3$ & $A_4$ & $A_5$ \\ 
\hline
$\log(\sigma_{\rm G})$&$4\leq G<7.5313$ & -4.31185&1.39006&
     -0.32577& 0.020446 & - & - \\
$\log(\sigma_{\rm G})$&$11.7230\leq G<12.7819$ & 467.32190 &-114.01847 & 
     9.19605 & -0.247111 & - & - \\
$\log(\sigma_{\rm G})$&$13.2107\leq G$ & -363.52090&113.42543&
     -14.20340&0.88228&-0.027177&
     0.00033255\\
\hline
Fitted uncertainty & Magnitude range & $R$ &$Q$&$P$ &&&\\ \cline{1-2}
\hline
$\ln[\log(\sigma_G)+4]$&$7.5313\leq G < 9.5060$ &    -12.16197&2.83670&
     -0.17781 & - & - & - \\
$\ln[\log(\sigma_G)+4]$&$9.5060\leq G < 10.7513$ &    -98.88558&19.38812&
     -0.95925 & - & - & - \\
$\ln[\log(\sigma_G)+4]$&$10.7513\leq G < 11.7230$ &    -241.49901&42.51819&
     -1.87684 & - & - & - \\
$\ln[\log(\sigma_G)+4]$&$12.7819\leq G < 13.2107$ &    -571.61430&88.12748&
     -3.402567 & - & - & - \\
\\
\hline
\end{tabular}
\label{tab:sigmafit}
\end{center}
\end{table*}


In order to improve the behaviour of the obtained predictions, for all fitted passbands, we restricted our fitting to the median of the observations as a function of the magnitude, and not considering the individual observations of all sources used to derive these medians. Table \ref{tab:sigmafit} shows the $G$ magnitude intervals and the corresponding coefficients for the different fitted laws. The resulting fitted laws in Table~\ref{tab:sigmafit} are plotted in Fig.~\ref{fig:sigma}.

\end{appendix}

\end{document}